\documentclass[11pt]{article}
\usepackage{jcapmod,parskip}
\usepackage{enumerate}
\usepackage{empheq}
\usepackage{feynmp}
\usepackage{booktabs}
\usepackage[english]{babel}
\usepackage{amsmath,amssymb,amsbsy,amstext, amsthm, simplewick}
\usepackage{graphicx}
\usepackage{amsfonts}
\usepackage{amssymb}
\usepackage{upgreek}
 \usepackage{exscale,relsize}
 \usepackage[makeroom]{cancel}
\usepackage{float}
\usepackage{framed}
\usepackage{array}
\newcolumntype{C}[1]{>{\centering\arraybackslash}p{#1}}
\RequirePackage{color}

\usepackage{colortbl}
\definecolor{rp}{cmyk}{0.2, 1, 0.6, 0}
\definecolor{green2}{cmyk}{0, 1, 0.5, 0}
\definecolor{lightgreen}{cmyk}{0.2, 0, 0.2, 0.2}
\definecolor{lightgray}{cmyk}{0.1,0.2,0,0.1}
\definecolor{lightgray2}{cmyk}{0.4,0.4,0,0.8}
\definecolor{black}{cmyk}{1.0,1.0,1.0,1.0}

\allowdisplaybreaks[1]

\usepackage{colortbl}
\definecolor{lightgreen}{cmyk}{0.2, 0, 0.2, 0.2}
\definecolor{lightgray}{cmyk}{0.1,0.2,0,0.1}
\definecolor{lightgray2}{cmyk}{0.1,0.1,0,0.1}

\makeatletter
\newlength{\apb@width}
\newcommand{\autoparbox}[2][c]{\settowidth{\apb@width}{#2}\parbox[#1]{\apb@width}{#2}}

\makeatother

\numberwithin{equation}{section}

\def\beq{\begin{equation}}
\def\eeq{\end{equation}}

\def\bea{\begin{eqnarray}}
\def\eea{\end{eqnarray}}

\def\be{\begin{equation}}
\def\ee{\end{equation}}
\def\ba{\begin{eqnarray}}
\def\ea{\end{eqnarray}}

\def\G{{\cal G}}

\def\0{{\boldsymbol 0}}
\def\k{{\boldsymbol{k}}}
\def\q{{\boldsymbol{q}}}
\def\p{{\boldsymbol{p}}}
\def\s{{\boldsymbol{s}}}

\def\x{{\boldsymbol{x}}}

\def\I{{\sf I}}
\def\J{{\sf J}}
\def\K{{\sf K}}

\DeclareRobustCommand{\SkipTocEntry}[4]{}

\begin{document}

\begin{titlepage}

\setcounter{page}{1} \baselineskip=15.5pt \thispagestyle{empty}

\bigskip\

\vspace{1cm}
\begin{center}
{\fontsize{20}{28}\selectfont  \sffamily \bfseries Cosmological Perturbation Theory Using the FFTLog: Formalism and Connection to QFT Loop Integrals
%A Flexible, Fast, and Fresh Approach to High-Order Calculations in Cosmological Perturbation Theory : Formalism
}
\end{center}

\vspace{0.5cm}

\begin{center}
{\fontsize{13}{30}\selectfont Marko Simonovi\'c,${}^{\rm a}$ Tobias Baldauf,${}^{\rm b}$ Matias Zaldarriaga,${}^{\rm a}$\\
John Joseph Carrasco${}^{\rm c}$ and Juna A. Kollmeier${}^{\rm d}$}
\end{center}

\begin{center}
\vskip 8pt
\textsl{${}^{\rm a}\;$School of Natural Sciences, Institute for Advanced Study, 1 Einstein Drive,\\ Princeton, NJ 08540, United States }\\ 
\vskip 2pt
\textsl{${}^{\rm b}\;$DAMTP, Center for Mathematical Sciences,
Wilberforce Road,\\
Cambridge, CB3 0WA, United Kingdom}\\
\vskip 2pt
\textsl{${}^{\rm c}\;$Institute of Theoretical Physics (IPhT), CEA-Saclay and University of Paris-Saclay\\
F-91191 Gif-sur-Yvette cedex, France}\\
\vskip 2pt
\textsl{${}^{\rm d}\;$The Observatories of the Carnegie Institution for Science, 813 Santa Barbara St,\\
Pasadena, CA 91101, United States}
\vskip 7pt
\end{center}

\vspace{1.2cm}
\hrule \vspace{0.3cm}
\noindent {\sffamily \bfseries Abstract} \\[0.1cm]
We present a new method for calculating loops in cosmological perturbation theory. This method is based on approximating a $\Lambda$CDM-like cosmology as a finite sum of complex power-law universes. The  decomposition is naturally achieved using an FFTLog algorithm. For power-law cosmologies, all loop integrals are formally equivalent to  loop integrals of massless quantum field theory. These integrals have analytic  solutions in terms of generalized hypergeometric functions. We provide explicit formulae for the one-loop and the two-loop power spectrum and the one-loop bispectrum. A chief advantage of our approach is that the difficult part of the calculation is cosmology independent, need be done only once, and can be recycled for any relevant predictions. Evaluation of standard loop diagrams then boils down to a simple matrix multiplication. We demonstrate the promise of this method for applications to higher multiplicity/loop correlation functions.
\vskip 10pt
\hrule

\vspace{0.6cm}
\end{titlepage}

\tableofcontents

\newpage
%%%%%%%%%%%%%%%%%%%%
\section{Introduction}
%%%%%%%%%%%%%%%%%%%%
Cosmological perturbation theory (PT)~\cite{Lifshitz:1946a, Fry:1983cj, Goroff:1986ep, Scoccimarro:1995if} offers a complementary approach to predicting and interrogating large-scale structure (LSS) observables in the weakly non-linear regime.  While many investigations in the literature are aimed at understanding and exploiting the highly non-linear regime (e.g. galaxy and cluster formation), it is clear that upcoming experiments (e.g. CMB-S4, DESI, CHIME) will also provide voluminous datasets probing  matter distribution on very large scales. As we continue in this ``era of precision cosmology", it is critical to exploit this investment of resources, to the fullest extent possible, in order to deliver on the promise of these upcoming surveys to measure cosmological parameters and potentially probe new physics.  The PT  approach, when recognized as a classical effective field theory~\cite{Baumann:2010tm,Carrasco:2012cv,Lewandowski:2014rca}, has a number of important advantages.  It converges to the correct answer for clustering statistics on large scales as more orders are included.  Its errors are parametrically controlled, so the typical size of the deviations from the correct answer can be easily estimated.  For the scales of relevance, it is rapidly computable compared to the necessarily large and high-resolution cosmological simulations required to attempt adequate comparison of theory and {\em large-scale} data.  Simulation boxes are computed with a single cosmology and thus many computations are required to investigate small changes in cosmological parameters, the effects of cosmic variance, and to perform consistency checks of scheme independence. PT works around these issues entirely.  Indeed it can serve as a spectacular large-scale (IR) complement to simulations, allowing them to focus their power in the incredibly non-linear smaller-scale (UV) regimes where they excel.
%\vskip 4pt

In the PT approach one treats dark matter and baryons as non-ideal self-gravitating fluids. At early times or on large scales these fluids are nearly homogeneous with small density fluctuations. This allows for the equations of motion to be solved perturbatively. The rigorous foundation of PT as an effective field theory of  large scale structure (EFTofLSS) was recently made~\cite{Baumann:2010tm,Carrasco:2012cv,Lewandowski:2014rca}, although many important results were known for a long time (for a review see~\cite{Bernardeau:2001qr}). One feature of these perturbative solutions is that they convolve initial density fields. Therefore observables, such as correlation functions of density-contrast, or {\em overdensity}, are efficiently written as momentum integrals over a certain number of initial power spectra. These integrals are refereed to as loop integrals. They admit a graphical organization which is why the atomic units of loop contributions to correlation functions are often referred to as loop diagrams.  Higher multiplicity/loop correlation calculations are critical not only to extending the scale of relevance of the analytic approach, but to break degeneracies and optimally extract cosmological parameters from the real data. 
%\vskip 4pt

While calculating loop integrals is a straightforward task in principle, the computational cost of exact solution becomes prohibitive for higher points (multiplicity) as well as higher loop order corrections. Higher multiplicity kernels quickly become  complicated and every loop brings additional three-dimensional integral. The linear power spectrum that appears in the integrand for real universe applications is known only as a numerical function which makes analytic solution of non-trivial integrals  impossible. Naive numerical integration, to desired precision, on the other hand, quickly becomes slow even with advanced Monte Carlo methods, due to the growth in dimensionality.  This poses a direct challenge to our ability to interrogate large datasets and one that merely more and faster computers will not address.
%\vskip 4pt

In order to simplify and speed up loop calculations we require new ideas, new strategies, to approach the problem. One inspiring idea, developed in~\cite{McEwen:2016fjn} and~\cite{Schmittfull:2016jsw}, is to use Fast Fourier Transform (FFT) for efficient evaluation of the one-loop power spectrum. After first ``deconvolving'' the lowest order PT solutions, and performing all angular integrals, the one-loop expressions reduce to a set of simple one-dimensional integrals that can be efficiently evaluated using FFT. Unfortunately, deconvolving higher order perturbative solutions and extending this approach to the one-loop bispectrum or the two-loop power spectrum proves to be challenging~\cite{Schmittfull:2016yqx}. 
%\vskip 4pt

In this paper we build on ideas of~\cite{McEwen:2016fjn,Schmittfull:2016jsw} but choose a slightly different strategy which allows us to go beyond the one-loop power spectrum. Let us briefly sketch the main idea behind our proposal. Prior to doing any integrals, the linear power spectrum is expanded as a superposition of ideal self-similar power-law cosmologies. This is naturally accomplished using FFT in $\log k$. Given some range of wavenumbers of interest, from $k_{\rm min}$ to $k_{\rm max}$, the approximation for the linear power spectrum with $N$ sampling points is \cite{Hamilton:1999uv,McEwen:2016fjn}
\be
\label{eq:fftlog}
\bar P_{\rm lin}(k_n) = \sum_{m=-N/2}^{m=N/2} c_m\,k_n^{\nu+i\eta_m} \;, 
\ee
where the coefficients $c_m$ and the frequencies $\eta_m$ are given by
\be
\label{eq:fftlogcoeff}
c_m = \frac 1N \sum_{l=0}^{N-1} P_{\rm lin}(k_l) \,k_l^{-\nu}k_{\rm min}^{-i\eta_m}\, e^{-2\pi i m l /N} \;, \quad \eta_m = \frac{2\pi m}{\log (k_{\rm max}/k_{\rm min})} \;.
\ee
Notice that the we denote the approximation for the linear power spectrum with $\bar P_{\rm lin}(k)$, while eq.~\eqref{eq:fftlogcoeff} uses the exact linear power spectrum $P_{\rm lin}(k)$ to calculate the coefficients $c_m$. We will keep using the same notation throughout the paper. The parameter $\nu$ is an arbitrary real number. As we will see, the simplest choice $\nu=0$ is insufficient in some applications, so we will use the more general form of the Fourier transform. In the terminology of \cite{McEwen:2016fjn} we call this $\nu$ parameter {\em bias}. Note that the powers in the power-law expansion are complex numbers. In practice, even a small number of power-laws, $\mathcal O(100)$, is enough to capture all features of the linear power spectrum including the BAO wiggles. One important thing to keep in mind is that the Fourier transform produces the power spectrum that is periodic in $\log k$. Therefore, we will take care to choose $k_{\rm min}$ and $k_{\rm max}$ such that we cover the range of scales where we actually care about the value of the power spectrum. In other words we are choosing the momentum range where the loop integrals have the most of the support. However, one always has to be careful about possible contributions particularly from high $k$ modes or short scales.  
%\vskip 4pt

Is this a limitation?  Absolutely not.  At the heart of the EFT understanding is the simple recognition that the PT idealized description of satisfying fluid-like equations of motion can only be valid at certain scales.  This is much the same as the hydrodynamic description of liquid water is only valid at certain scales. Attempting to integrate this approximation over scales outside of its validity introduces non-parametrically controlled errors.  Instead the information in the linear approximation must be supplemented by small-scale UV physics. This data is encoded in physical parameters like speed of sound or viscosity -- potentially any dimensionally consistent operators. Such EFT parameters serve two roles.  They must eat up any cutoff-dependence, by definition non-physical, and they must accurately represent the integrating out of any small-scale degrees of freedom.  So baked into the framework that places PT on a rigorous footing is the realization that any integrals of the linear approximation should only be performed over a range of scales consistent with its validity.
%\vskip 4pt

Notice that the decomposition \eqref{eq:fftlog} reduces the evaluation of a loop diagram for an arbitrary cosmology to evaluation of the same diagram for a set of different power-law universes with numerical coefficients. Power-law momentum integrals {\em can} be done analytically. The final answer is a sum of familiar special functions which are straightforward to evaluate. In the simplest case of the one-loop power spectrum, the momentum integral for a power-law universe can be expressed entirely in terms of gamma functions \cite{Scoccimarro:1996jy,Pajer:2013jj}. Looking at higher order correlators an interesting pattern emerges. The perturbation theory kernels can always be written such that the general form of loop integrals in a power-law cosmology is formally identical to the one of a massless Quantum Field Theory (QFT) with cubic interactions.\footnote{More precisely, it is a QFT in three dimensions with the Euclidean signature.} This is just a formal relationship, but it should prove rather useful. Many results recently developed in the theory of scattering amplitudes can be applied to LSS correlation functions as well. Some steps in this direction have already been made for the one-loop bispectrum in \cite{Scoccimarro:1996jy}. 
%\vskip 4pt

In this work we derive formulas for the one-loop bispectrum and the two-loop power spectrum in power-law cosmologies which are suitable for effecient numerical evaluation. Generically, the higher multiplicity/loop correlation functions are expressed in terms of the generalized hypergeometric functions. One thing to keep in mind is that the powers $\nu+i\eta_m$ are complex and one has to be careful about the analytic continuation of all results to the entire complex plane.
%\vskip 4pt

Before diving into the details, let us comment on one important virtue of our method. The decomposition~\eqref{eq:fftlog} is useful because it separates the cosmology dependent portion, encoded entirely in the coefficients $c_m$, from the loop calculations which have been reduced to that of much more tractable, ideal, cosmologies. This means that for the fixed value of bias $\nu$, the momentum range $(k_{\rm min},k_{\rm max})$ and the number of sampling points $N$ the difficult part of the calculation which involves momentum integrals can be done only once, saved as a table of numbers and then used for any cosmology. As we will see, the evaluation of the contribution of loop diagrams reduces to a simple (small) matrix multiplication and it is very fast. This opens up the possibility of using our method in Markov chain Monte Carlo parameter estimation.
%\vskip 4pt

In the rest of the paper we focus on three examples: the one-loop power spectrum, the one-loop bispectrum and the two-loop power spectrum. We will present our calculations in detail and for one-loop diagrams compare them with the standard numerical results. We leave a detailed comparison with the numerical two-loop power spectrum for future work. A {\sf Mathematica} notebook used to produce plots is available as an auxiliary file associated with the preprint of the paper on {\sf arXiv}. 

\section{One-loop Power Spectrum}
\begin{figure}[h]
\begin{center}
\includegraphics[width=0.7\textwidth]{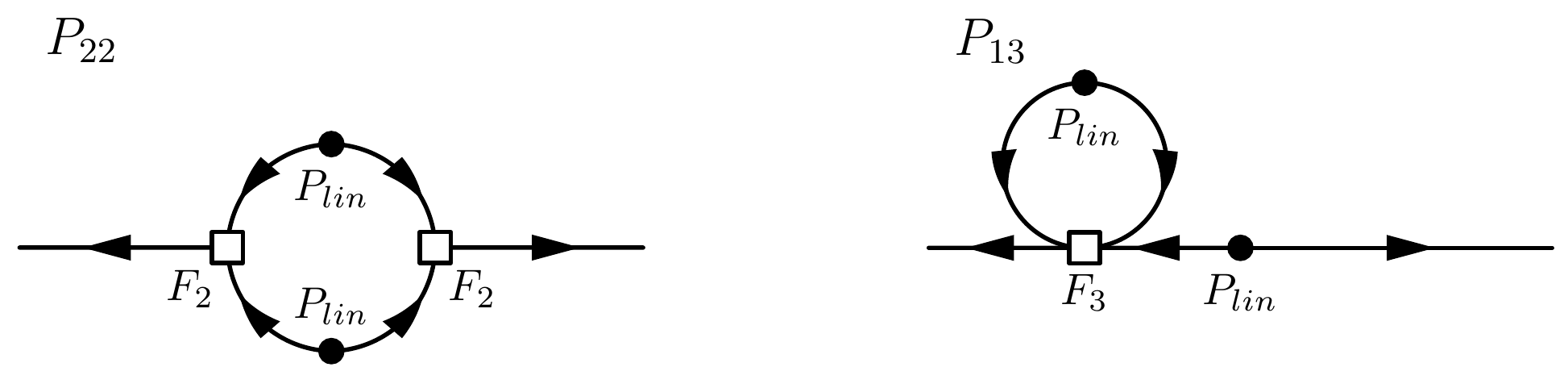}
\end{center}
\caption{Diagrammatic representation of two contributions to the one-loop power spectrum.}
\label{fig:diagrams1LP}
\end{figure}

Let us first consider the simplest case---the one-loop power spectrum. In perturbation theory there are two different one-loop contributions. Using the usual approximation in which the time dependence is separated from $k$ dependence (for a review see~\cite{Bernardeau:2001qr}), the one-loop power spectrum reads
\be
\label{eq:P1loop}
P_{\rm 1-loop}(k,\tau) = D^4(\tau)[P_{22}(k)+P_{13}(k)] \;,
\ee
where $\tau$  is conformal time, $D(\tau)$ is the growth factor for matter fluctuations and the two terms in the square brackets are given by
\be
\label{eq:P22def}
P_{22}(k) = 2 \int_{\q} F_{2}^2(\q, \k - \q) P_{\rm lin}(q) P_{\rm lin}( |\k- \q| ) \;,
\ee
\be
\label{eq:P13def}
P_{13}(k) = 6P_{\rm lin}(k) \int_{\q} F_3(\q, -\q, \k) P_{\rm lin}(q) \;,
\ee
where $\int_{\q}\equiv\int\frac{d^3q}{(2\pi)^3}$. Diagrammatic representation of these two contributions is shown in Fig.~\ref{fig:diagrams1LP}. The explicit form of kernels $F_n$ can be calculated using well-known recursion relations~\cite{Bernardeau:2001qr}. One important point is that it is always possible to expand kernels in~\eqref{eq:P22def} and~\eqref{eq:P13def} in integer powers of $k^2$, $q^2$ and $|\k-\q|^2$. For example,
\begin{align}
\label{eq:F2expansion}
F_2(\q,\k-\q) & = \frac{5}{14} +\frac{3 k^2}{28 q^2} +\frac{3 k^2}{28 |\k-\q|^2}-\frac{5 q^2}{28 |\k-\q|^2}-\frac{5 |\k-\q|^2}{28 q^2} +\frac{k^4}{14 |\k-\q|^2 q^2}.
\end{align}
A similar expression can be found for $F_3(\q,-\q,\k)$.\footnote{In the expansion of $F_3(\q,-\q,\k)$ some terms contain $|\k+\q|^2$. Given that the kernels are always integrated over $\q$, one is allowed to do the following change of coordinates $\q\to-\q$ and bring these terms to the same form as in~\eqref{eq:F2expansion}} If we further decompose $P_{\rm lin}(k)$ in power laws using~\eqref{eq:fftlog}, the one-loop power spectrum becomes a sum of simple momentum integrals of the following form
\be
\label{eq:Iintdef}
\int_{\q} \frac{1}{q^{2\nu_1}|\k-\q|^{2\nu_2} } \equiv k^{3-2\nu_{12}}\I(\nu_1,\nu_2) \;,
\ee
where $\nu_1$ and $\nu_2$ are in general complex numbers. 
%\vskip 4pt

As we already mentioned, the form of the integral is identical to the one-loop massless two point function in QFT. The only difference is that in this case the powers of the ``propagators" are complex numbers rather than integers. Still, the unknown dimensionless function $\I(\nu_1,\nu_2)$ can be easily calculated using the standard technique with Feynman parameters. The result is a well known expression~\cite{Scoccimarro:1996jy,Pajer:2013jj}
\be
\label{eq:Iint}
\I(\nu_1,\nu_2) = \frac{1}{8\pi^{3/2}} \frac{\Gamma(\tfrac32 - \nu_1) \Gamma(\tfrac32 - \nu_2) \Gamma(\nu_{12} - \tfrac32)}{\Gamma(\nu_1) \Gamma(\nu_2) \Gamma(3 - \nu_{12})} \;,
\ee
were $\nu_{12}=\nu_1+\nu_2$ (throughout the paper we adopt the following notation $\nu_{1...n}\equiv \nu_1+\cdots +\nu_n$). Notice that, thanks to the analytic continuation, $\I(\nu_1,\nu_2)$ gives a finite answer even for the values of parameters for which the integral is formally divergent. In practice, breaking the loop calculation into many pieces can lead to some divergent terms. However, as long as the {\em total} sum is well defined and finite, for at least some power-law cosmology $P_{\rm lin}(k)\sim k^\nu$, by analytic continuation it is guaranteed that eq.~\eqref{eq:Iint} gives the correct answer. 
%\vskip 4pt

Sometimes the condition that the integral at hand is convergent for at least some power-law power spectrum cannot be met, and one has to use eq.~\eqref{eq:Iint} with some care. For example, the function $\I(\nu_1,\nu_2)$ vanishes if one of the arguments is zero (or a negative integer). Applying~\eqref{eq:Iint} blindly would lead in these cases to paradoxical results. For instance, after power-law decomposition of the linear power spectrum, eq.~\eqref{eq:Iint} would imply 
\be
\label{eq:paradox}
\int_{0}^\infty dq\,P_{\rm lin}(q) = 0 \;,
\ee
which is obviously the wrong answer. This is a consequence of the well known statement that in dimensional regularization all power-law divergences vanish: $\int_\q q^\nu = 0$.\footnote{More precisely, this integral is related to a delta function \cite{Gorishnii:1984te}. A  change of coordinates relates
\be
\int_\q \frac{1}{q^{3+2\nu_1}} = \frac{i}{2\pi^2} \delta(\nu_1) \;.
\ee
To get the consistent results one can use this equation. In practice, there is a much simpler way, as described in the main text. 
}
%\vskip 4pt 

Similar issues can appear in calculating loop diagrams. Luckily, for a $\Lambda$CDM-like cosmology, they can be always easily fixed. Let us imagine that the integral we are interested in is divergent for a given bias $\nu$. Then, if the integral diverges in the UV(IR), one has to find the UV(IR) limit of the integrand. This can be easily done fixing all external momenta and sending the loop momentum to infinity(zero). This limit always has the form of eq.~\eqref{eq:paradox} and it would be set to zero by dimensional regularization. Therefore, to get the correct answer, one simply has to add the UV(IR) contribution by hand. In the following sections we will give more details for each specific case we consider.  
%\vskip 4pt

Let us also point out that all UV divergences have a well defined momentum dependence. This momentum dependence is the same as for the counterterms in the EFTofLSS. Therefore, one can proceed without explicitly adding the UV-dependent terms to the loop calculation. The only effect of this choice is to change the usual values of the counterterms. In this sense we can say that eq.~\eqref{eq:Iint} calculates only the ``finite'' part of the loop integral. As expected, the counterterms absorb all UV-dependent pieces. 
\vskip 10pt

\subsection{Symmetries of $\I(\nu_1,\nu_2)$ and Recursion Relations}
Before moving on to applications, it is instructive to take a closer look at symmetries of the integral in~\eqref{eq:Iintdef}. For the one-loop power spectrum this is a straightforward exercise. We use it to introduce some notation and derive a couple of results that will be very useful in more complicated cases, such as the one-loop bispectrum or the two-loop power spectrum. 
%\vskip 4pt

The most obvious symmetry of the integral~\eqref{eq:Iintdef} is invariance under the shift $\q \to\k-\q$. Consequently, the function $\I(\nu_1,\nu_2)$ is symmetric in $\nu_1$ and $\nu_2$
\be
\I(\nu_1,\nu_2) = \I(\nu_2,\nu_1) \;.
\ee
As we will see later, there are similar transformations for higher multiplicity/order diagrams and they always lead to some permutation of parameters $\nu_i$. We are going to call these kind of identities {\em translation} formulas, because they are derived using translations in momentum space. 
%\vskip 4pt

The integral~\eqref{eq:Iintdef} preserves its form under rescaling of momenta, but this does not lead to any non-trivial condition on $\I(\nu_1,\nu_2)$. However, a more complicated rescaling, such as inversion of momenta, does lead to interesting results. For simplicity, let us choose $\k$ to be a unit vector: $\k=\hat\k$, $\hat\k^2=1$. Under an inversion $\q\to \q/q^2$ different factors in the integrand transform in the following way
\be
\label{eq:Iinvtransf}
d^3q\;\to\; \frac{d^3q}{q^6}\;, \quad q^2\;\to\;q^{-2}\;, \quad |\hat\k-\q|^2\;\to\;\frac{|\hat\k-\q|^2}{q^2} \;.
\ee  
Using  these transformations we can write
\be
\I(\nu_1,\nu_2) = \int_{\q} \frac{1}{q^{2\nu_1}|\hat\k-\q|^{2\nu_2}} = \int_{\q} \frac{1}{q^{2(3-\nu_{12})}|\hat\k-\q|^{2\nu_2} } \;,
\ee
which immediately implies the {\em inversion} formula
\be
\I(\nu_1,\nu_2) = \I(3-\nu_{12},\nu_2)\;.
\ee
%\vskip 4pt

There is one more method to find nontrivial identities for massless loop integrals which is based on the following relation between real and momentum space
\be
\label{eq:duality}
\frac1{q^{2\nu}} = \frac{\Gamma(\tfrac32-\nu)}{\Gamma(\nu)} \pi^{-3/2} 2^{-2\nu} \int_{\x} \frac1{x^{3-2\nu}} e^{-i\q\cdot\x} \;,
\ee
where $\int_{\x}\equiv \int d^3x$. Let us illustrate the main idea behind this method. The starting point is to close all external lines in order to form additional loops. This is equivalent to integrating over all external momenta. For example, for the one-loop power spectrum we can start from
\be
\int_{\k}\frac1{k^3}\,\I(\nu_1,\nu_2) = \int_{\k,\q} \frac{1}{q^{2\nu_1}|\k-\q|^{2\nu_2}k^{2(3-\nu_{12})}} \,.
\ee
Notice that this expression has a form of a two-loop vacuum diagram. We have chosen to multiply $\I(\nu_1,\nu_2)$ with the $1/k^3$ factor such that the whole integral is dimensionless. On the l.h.s.~the function $\I(\nu_1,\nu_2)$ does not depend on $k$ and the integral trivially reduces to
\be
\int_{\k}\frac1{k^3}\,\I(\nu_1,\nu_2) = \I(\nu_1,\nu_2) \frac1{(2\pi)^3}\int\frac{d^3k}{k^3} \;.
\ee
On the r.h.s.~we can use~\eqref{eq:duality} and integrate over $\q$ and $\k$. The momentum integrals lead to two delta functions which can be used to do two integrals in real space. At the end of the day, we are left with the following expression
\be
\I(\nu_1,\nu_2) \frac1{(2\pi)^3}\int\frac{d^3k}{k^3} = \frac{1}{64\pi^{9/2}} \frac{\Gamma(\tfrac32 - \nu_1) \Gamma(\tfrac32 - \nu_2) \Gamma(\nu_{12} - \tfrac32)}{\Gamma(\nu_1) \Gamma(\nu_2) \Gamma(3 - \nu_{12})} \int\frac{d^3x}{x^3} \;,
\ee
from which result~\eqref{eq:Iint} immediately follows. As we can see, the one-loop two-point function is simple enough that relation~\eqref{eq:duality} is sufficient to fix its form. For higher multiplicity/loop correlation functions this is not the case. The reason is that in those cases the real space integrals are not trivial anymore. However, it turns out that they alway have the same structure as the original momentum integrals. The only difference is that the parameters are shifted:~$\nu_i\to \tilde\nu_i\equiv \tfrac 32-\nu_i$. It follows that there is always an identity which relates two functions with parameters $\nu_i$ and $\tilde\nu_i$. We will refer to these identities as {\em reflection} formulas.
%\vskip 4pt 

Finally, let us present recursion relations which connect functions whose parameters differ by an integer. These relations can be always derived using the fact that the integral of a total derivative vanishes. For instance, 
\be
\int_\q \frac{\partial}{\partial q_i}\left( \frac{q_i}{q^{2\nu_1}|\k-\q|^{2\nu_2}} \right) = 0 \;.
\ee
Expanding the derivative we find
\be
\label{eq:Irecursion}
(3-2\nu_1-\nu_2)\I(\nu_1,\nu_2)+\nu_2[\I(\nu_1,\nu_2+1)-\I(\nu_1-1,\nu_2+1)]=0 \;,
\ee
and a similar relation in which $\nu_1$ and $\nu_2$ are exchanged. The importance of these identities is that they relate different terms in the expansion of kernels, such as~\eqref{eq:F2expansion}. As we will see in the explicit calculations of $P_{22}$ and $P_{13}$ diagrams, thanks to the recursion relations all terms in the expansion of the kernels (for fixed $\nu_1$ and $\nu_2$) can be evaluated using a single function $\I(\nu_1,\nu_2)$.  
\vskip 10pt

\subsection{Numerical Evaluation of the One-loop Power Spectrum}
In this section we will apply eq.~\eqref{eq:Iint} to the calculation of the one-loop power spectrum. We will first separately discuss $P_{22}$ and $P_{13}$ diagrams (see Fig.~\ref{fig:diagrams1LP}).
%\vskip 4pt

\noindent
{\em $P_{22}$ diagram.---}Let us begin by reviewing some properties of the $P_{22}$ diagram in a cosmology with $P_{\rm lin}(k)\sim k^\nu$. In particular, we are interested in finding the powers $\nu$ for which the integral is convergent. In order to do that we have to find the asymptotic form of the integrand in the UV and the IR regime. The behavior of the $F_2$ kernel in these two limits is 
\be
F_2(\q,\k-\q) \to \frac kq \;, \qquad q\to 0 \;,
\ee
\be
F_2(\q,\k-\q) \to \frac{k^2}{q^2} \;, \qquad q\to \infty \;.
\ee
It follows that $P_{22}$ diagram is convergent if $-1<\nu<1/2$. If we choose bias in FFT to be in this range, then the integral in $P_{22}$ is finite for each term in the sum~\eqref{eq:fftlog} and using~\eqref{eq:Iint} we are guaranteed to get the same answer as with the usual numerical evaluation. 
%\vskip 4pt

\begin{figure}[h]
\includegraphics[width=0.5\textwidth]{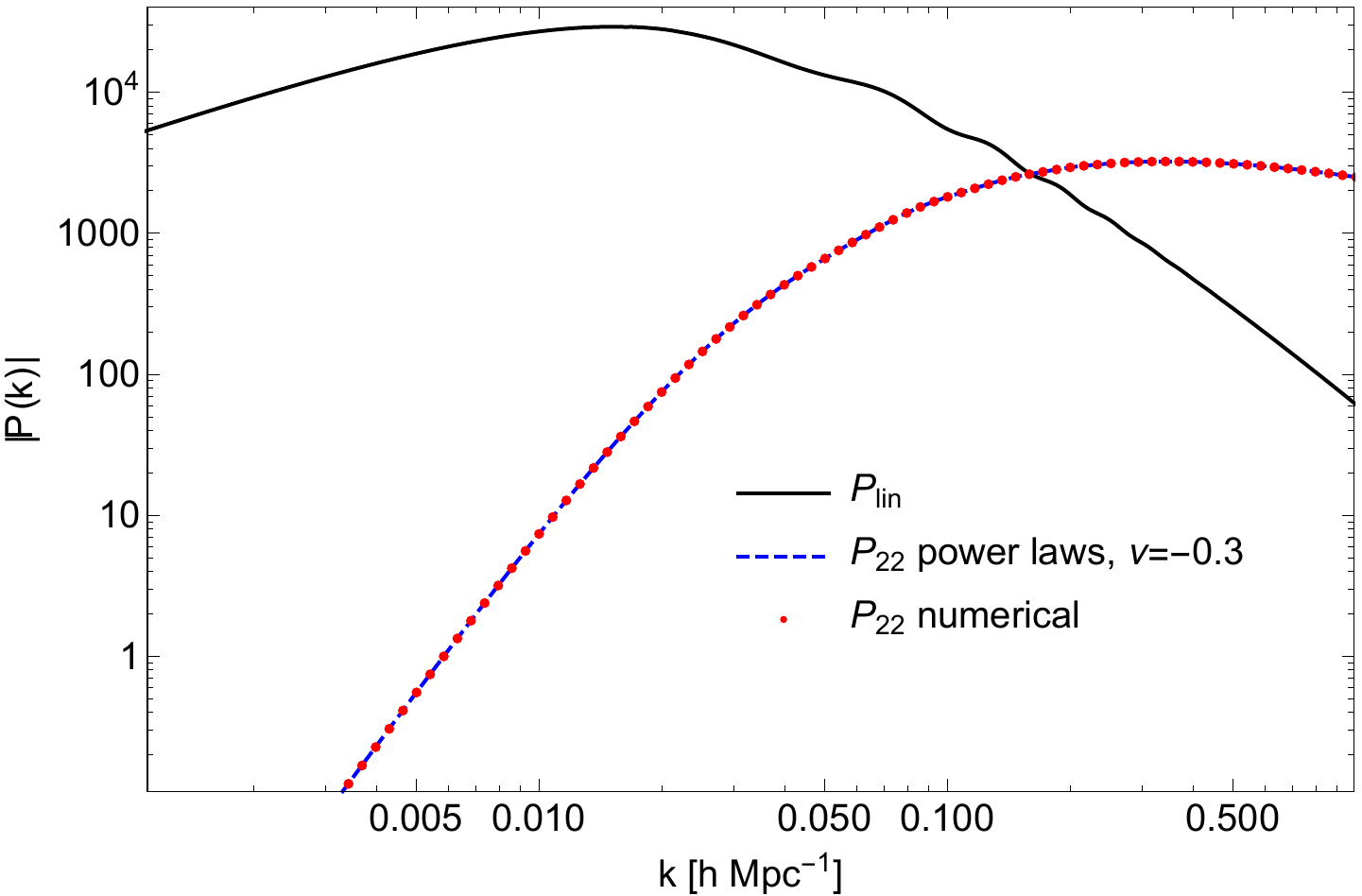}
\includegraphics[width=0.5\textwidth]{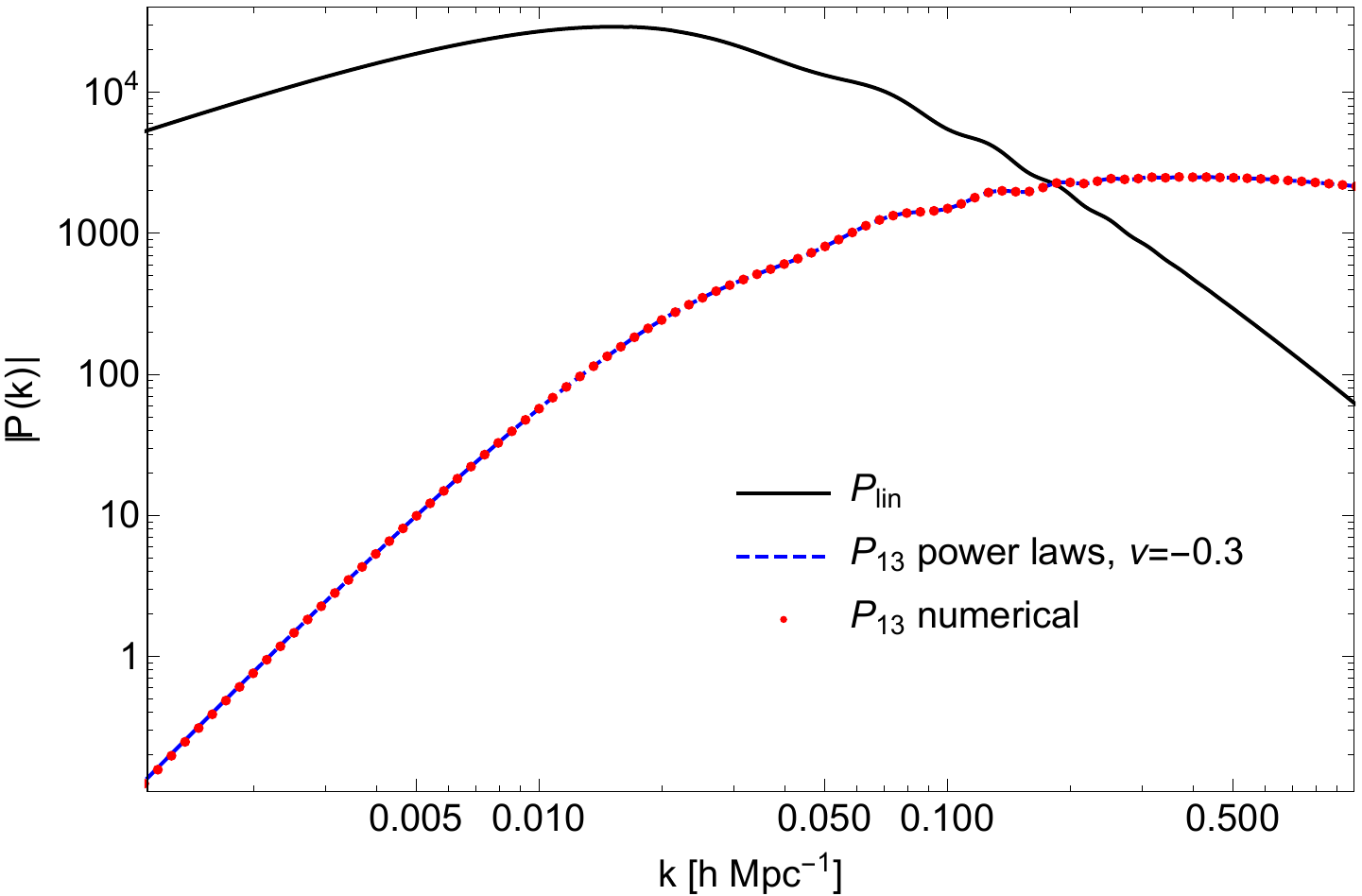}
\caption{Two contributions to the one-loop power spectrum calculated using direct numerical integration and eq.~\eqref{eq:P22powers1} and eq.~\eqref{eq:P13powers1} as described in the main text. Both plots are produced using $\nu=-0.3$, $N=150$, $k_{\rm min}=10^{-5}\,h{\rm Mpc}^{-1}$ and $k_{\rm max}=5\,h{\rm Mpc}^{-1}$. For these values of parameters the sum of two terms differs from the numerical one-loop power spectrum  by less than $0.1\%$ at all scales.}\label{fig:p22p13correct}
\end{figure}

Before turning to results, let us write the explicit formula for $P_{22}$ diagram. Using~\eqref{eq:fftlog} and~\eqref{eq:F2expansion} we can write the approximation to the $P_{22}$ diagram in the following way 
\be
\label{eq:P22powers}
\bar P_{22}(k) = 2 \sum_{m_1,m_2} c_{m_1}\,c_{m_2} \sum_{n_1,n_2} f_{22}(n_1,n_2)\,k^{-2(n_1+n_2)} \int_{\q} \frac{1}{q^{2\nu_1-2n_1}|\k-\q|^{2\nu_2-2n_2} } \;.
\ee
In this expression $c_{m_1}$ and $c_{m_2}$ are the coefficients in \eqref{eq:fftlogcoeff} and $n_1$ and $n_2$ are integer powers of $q^2$ and $|\k-\q|^2$ in the expansion of $F_2^2(\q,\k-\q)$. Corresponding rational coefficients in this expansion are labeled by $f_{22}(n_1,n_2)$ and they can be read off from~\eqref{eq:F2expansion}. The complex numbers $\nu_1$ and $\nu_2$ are given by
\be
\label{eq:formnui}
\nu_1 = -\tfrac12(\nu+i\eta_{m_1}) \quad {\rm and } \quad \nu_2 = -\tfrac12(\nu+i\eta_{m_2}) \;.
\ee
Using the solution for the momentum integral, expression~\eqref{eq:P22powers} can be further simplified and written in the following way
\be
\label{eq:P22powers1}
\bar P_{22}(k) = k^{3} \sum_{m_1,m_2} c_{m_1}k^{-2\nu_1}\cdot M_{22}(\nu_1,\nu_2) \cdot c_{m_2}k^{-2\nu_2} \;,
\ee
where the matrix $M_{22}(\nu_1,\nu_2)$ is given by
\be
\label{eq:matrixM22}
M_{22}(\nu_1,\nu_2)= \frac{(\tfrac32 - \nu_{12}) (\tfrac12 - \nu_{12}) [\nu_1\nu_2 (98\nu_{12}^2 - 14\nu_{12}+ 36 ) - 91 \nu_{12}^2 + 3\nu_{12} + 58]}{196\, \nu_1 (1+\nu_1) (\tfrac12 - \nu_1)\,\nu_2 (1+\nu_2) (\tfrac12-\nu_2)} \,\I(\nu_1,\nu_2).
\ee
As we already pointed out, only a single function $\I(\nu_1,\nu_2)$ is sufficient to calculate the full diagram. Thanks to the recursion relations~\eqref{eq:Irecursion}, all terms from the expansion of $F_2$ kernels are encoded in the $\nu$-dependent prefactor in matrix $M_{22}(\nu_1,\nu_2)$.
%\vskip 4pt 

One can use eq.~\eqref{eq:P22powers1} to calculate the $P_{22}$ diagram. The result is shown in Fig.~\ref{fig:p22p13correct}. As expected, the agreement with the usual numerical integration is excellent. An important thing to notice is that the only cosmology dependence in~\eqref{eq:P22powers1} is in the coefficients $c_m$. For a given number of sampling points $N$, bias $\nu$ and $k_{\rm min}$ and $k_{\rm max}$, the matrix $M_{22}(\nu_1,\nu_2)$ is fixed. This means that it can be calculated only once and saved as a table of numbers. The evaluation of the $P_{22}$ diagram for an arbitrary cosmology then boils down to doing one FFT to determine coefficients $c_m$, calculating a vector $c_mk^{-2\nu}$ for each $k$ and a simple matrix multiplication~\eqref{eq:P22powers1}.
%\vskip 4pt

So far we restricted ourselves to biases in the range $-1<\nu<\tfrac 12$. Outside this range one has to be more careful because the integrals are not convergent anymore and eq.~\eqref{eq:Iint} is not guaranteed to give the correct answer. For example, for biases $\nu<-1$ the integrals are divergent in the IR. The leading piece of the $P_{22}$ diagram in this limit can be calculated by fixing $k$ and sending $\q\to 0$ in the integrand. The result is
\be
\label{eq:p22IR}
P_{22}^{\rm IR} (k) = P_{\rm lin}(k) k^2 \sigma_v^2 \;,
\ee
where $\sigma_v^2\equiv\tfrac{1}{6\pi^2}\int_0^\infty dq\,P_{\rm lin}(q)$. The integral in $\sigma_v^2$ would be set to zero by eq.~\eqref{eq:Iint} and missed in the final answer. Therefore, to get a correct result, one simply has to add $P_{22}^{\rm IR} (k)$ to eq.~\eqref{eq:P22powers1} at the end of the calculation. Notice that we have kept only the leading IR divergence, which is enough for biases in the range $-3<\nu<-1$. If the bias was even smaller, one would have to keep track of sufficient number of subleading IR divergences. Similarly, when $\tfrac12<\nu<\tfrac32$, the leading UV divergence that has to be added on the r.h.s.~of eq.~\eqref{eq:P22powers1} to get the correct result is 
\be
\label{eq:p22UV}
P_{22}^{\rm UV} (k) = \frac9{196\pi^2}k^4 \int_0^\infty dq \frac{P_{\rm lin}(q)}{q^2}\;. 
\ee % Fixed 2nd eq:p22IR -> eq:p22UV 
However, for very high or low values of biases, the momentum range $(k_{\rm min},k_{\rm max})$ has to be very wide for the integrals to converge to correct values. This implies a large number of frequencies and it is not practical.
%\vskip 4pt

\noindent
{\em $P_{13}$ diagram.---}The asymptotic behavior of the $F_3$ kernel in the UV and the IR limit is the same 
\be
F_3(\q, -\q, \k) \to  \frac{k^2}{q^2}  \;, \qquad q\to 0 \quad {\rm or} \quad q\to \infty\;.
\ee
Consequently, $P_{13}$ diagram is divergent in the UV for $\nu>-1$ and divergent in the IR for $\nu<-1$. In other words, $P_{13}$ diagram is never finite in a power-law cosmology. For $\nu>-1$, the only possible mismatch between the true answer and eq.~\eqref{eq:Iint} comes from the UV part of the integral. For fixed $k$ and taking the limit $\q\to\infty$  
\be
\label{eq:p13UV}
P_{13}^{\rm UV} (k) = -\frac{61}{105} P_{\rm lin}(q) k^2 \sigma_v^2\;.
\ee
On the other hand, for $\nu<-1$, the possible error comes from the IR limit\footnote{Notice that $P_{13}$ diagram has two IR divergences $\q\to0$ and $\q\to\k$, which are combined in a single expression.}
\be
\label{eq:p13IR}
P_{13}^{\rm IR} (k) = -P_{\rm lin}(k) k^2 \sigma_v^2 \;.
\ee
As we already mentioned, in dimensional regularization this type of the integrals would be set to zero by~\eqref{eq:Iint}. This implies that to get the correct values for $P_{13}$, depending on the choice of bias, we have to add either the UV or IR term by hand. Notice that here we are writing down only the leading IR and UV parts of the $P_{13}$ diagram. For biases $\nu<-3$ or $\nu>1$, one would have to include the corresponding subleading terms as well.
%\vskip 4pt 
\begin{figure}[h]
\includegraphics[width=0.5\textwidth]{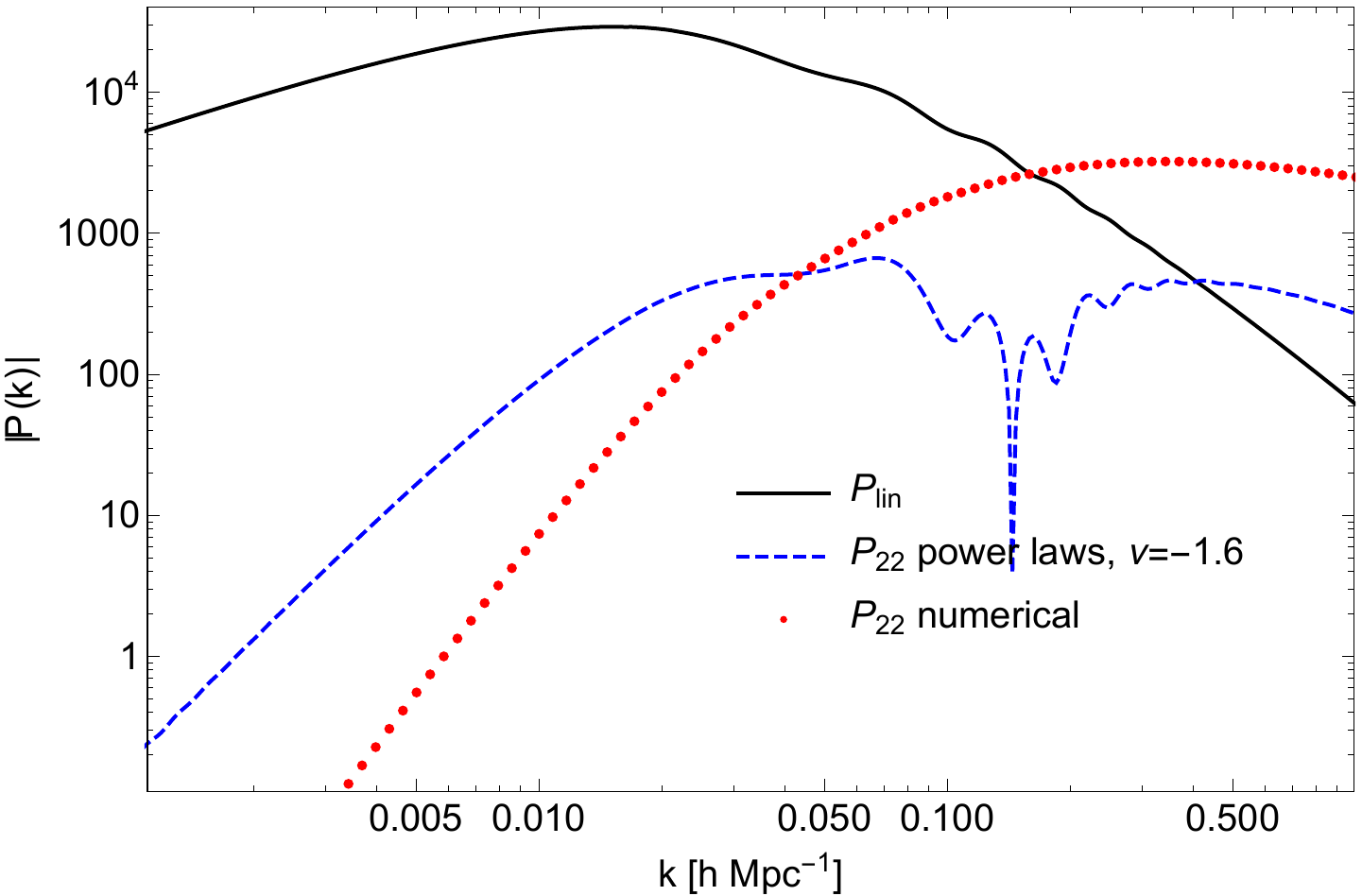}
\includegraphics[width=0.5\textwidth]{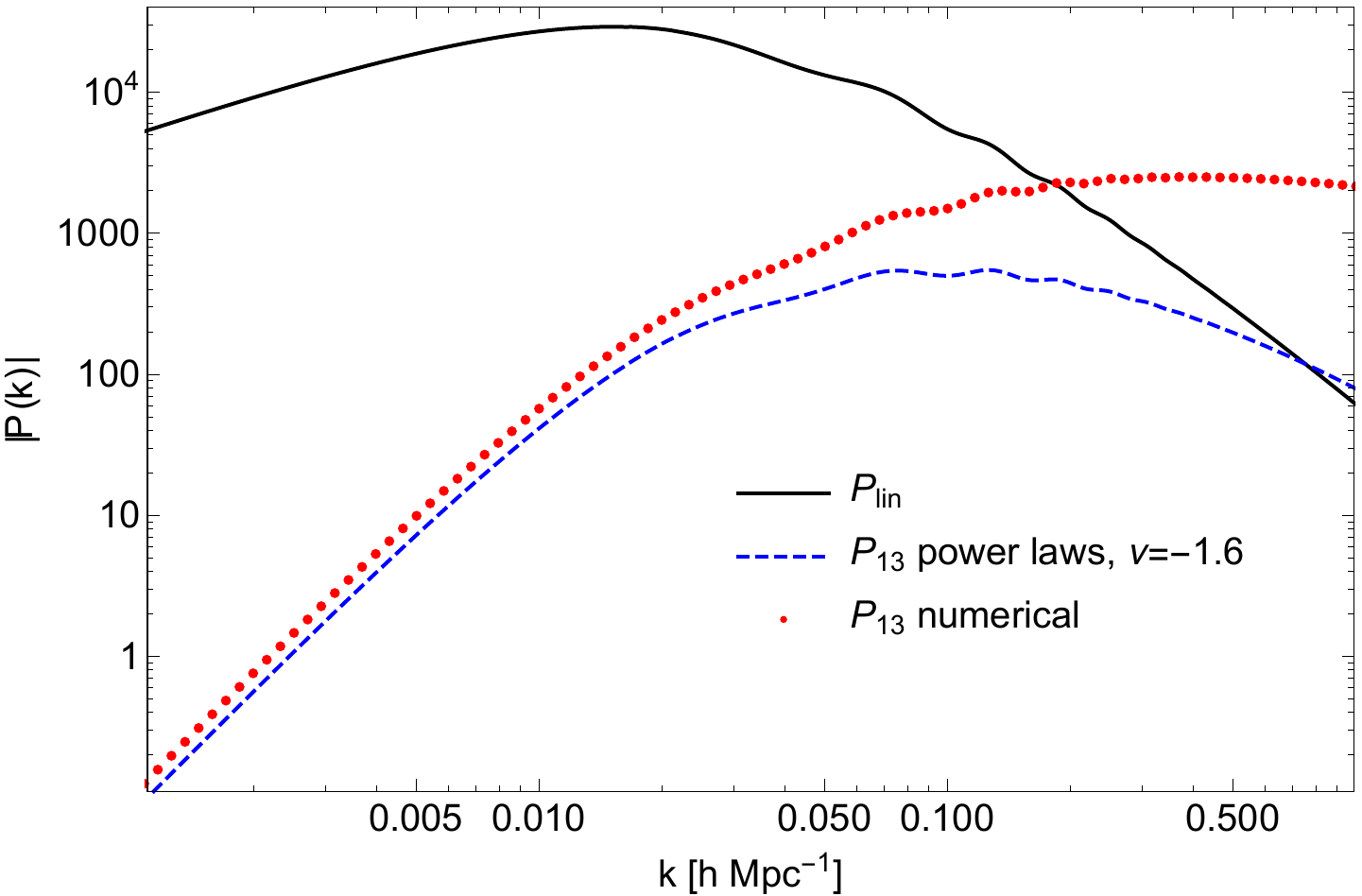}
\caption{Two contributions to the one-loop power spectrum calculated using direct numerical integration and eq.~\eqref{eq:P22powers1} and eq.~\eqref{eq:P13powers1}. Both plots are produced using $\nu=-1.6$, $N=150$, $k_{\rm min}=3\cdot 10^{-4}\,h{\rm Mpc}^{-1}$ and $k_{\rm max}=180\,h{\rm Mpc}^{-1}$. For this value of bias both $P_{22}$ and $P_{13}$ are very different from their standard values.}
\label{fig:p22p13wrong}
\end{figure}

Let us see how the formulas above work in practice. With the same notation as for the $P_{22}$ diagram, we can write 
\be
\label{eq:P13powers}
\bar P_{13}(k) = 6P_{\rm lin}(k) \sum_{m_1} c_{m_1} \sum_{n_1,n_2} f_{13}(n_1,n_2)\,k^{-2(n_1+n_2)} \int_{\q} \frac{1}{q^{2\nu_1-2n_1}|\k-\q|^{-2n_2} } \;.
\ee
Solving the momentum integral, this expression can be further simplified 
\be
\label{eq:P13powers1}
\bar P_{13}(k) = k^{3} P_{\rm lin}(k) \sum_{m_1} c_{m_1}k^{-2\nu_1} \cdot M_{13}(\nu_1)\;,
\ee
where the vector $M_{13}(\nu_1)$ is given by
\be
M_{13}(\nu_1) = \frac{1 + 9 \nu_1}4 \frac{\tan(\nu_1 \pi)}{28 \pi (\nu_1 + 1) \nu_1 (\nu_1 - 1) (\nu_1 - 2) (\nu_1 - 3)}\;.
\ee
Notice that to eq.~\eqref{eq:P13powers1} one has to add the UV or the IR part of the integral. For example, for $\nu>-1$, we plot the result in Fig.~\ref{fig:p22p13correct}. As expected, once $P_{13}^{\rm UV}(k)$ is added to eq.~\eqref{eq:P13powers1}, the agreement with the usual numerical result is excellent.
%\vskip 4pt

\begin{figure}[h]
\begin{center}
\includegraphics[width=0.7\textwidth]{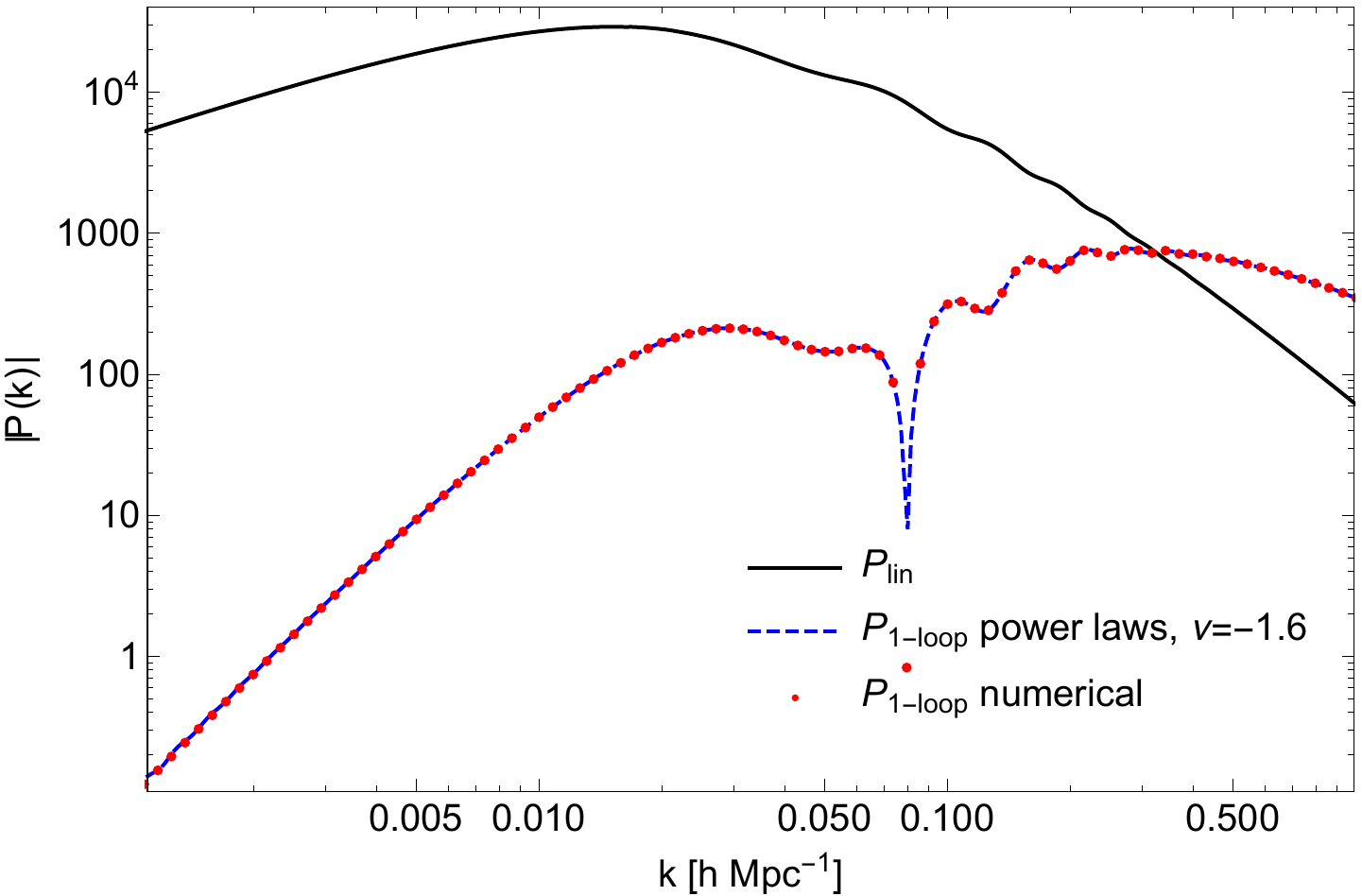}
\end{center}
\caption{The full one-loop power spectrum calculated summing up contributions from Fig.~\ref{fig:p22p13wrong}. }
\label{fig:p1loop}
\end{figure}

\noindent
{\em The full one-loop power spectrum.---}So far we were trying to reproduce the usual numerical results for separate pieces of the one-loop power spectrum. However, only their sum is a well defined observable. Thanks to the Equivalence Principle the IR divergences cancel and the {\em total} one-loop power spectrum is convergent for the range of power laws $-3<\nu<-1$ \cite{Pajer:2013jj}. This means that with the choice of bias in this range, the formulas above should lead to the correct answer for $P_{\rm 1-loop}(k)$, without having to deal with the IR divergences explicitly. In Fig.~\ref{fig:p1loop} we plot the one-loop power spectrum calculated in this way and show that our method indeed agrees with the usual numerical result. As expected, the separate terms $P_{13}$ and $P_{22}$ are rather different from their usual values (see Fig.~\ref{fig:p22p13wrong}). However, the ``mistake'' that eq.~\eqref{eq:Iint} makes in assigning some finite values to divergent integrals has to cancel between the two contributions in the same way the IR divergences cancel. Indeed, the IR limit of the $P_{22}$ diagram is exactly the same as the IR limit of $P_{13}$ diagram, but with the opposite sign (see eq.~\eqref{eq:p22IR} and eq.~\eqref{eq:p13IR}).
%\vskip 4pt

In conclusion, the one-loop power spectrum can be easily calculated using decomposition~\eqref{eq:fftlog}. For practical applications, the most optimal choice of bias is in the range close to zero $-0.5<\nu<0$ because it requires the least number of frequencies to reproduce the linear power spectrum on relevant scales. For this range the $P_{22}$ diagram can be evaluated directly applying our method. To get the correct $P_{13}$ diagram, one has to add $P_{13}^{\rm UV}$ term to the r.h.s~of eq.~\eqref{eq:P13powers1}. 
\vskip 10pt

\subsection{One-loop Power Spectrum of Biased Tracers}
The method described above can be also applied to the one-loop power spectrum of biased tracers \cite{McDonald:2009dh,Senatore:2014eva,Assassi:2014fva,Mirbabayi:2014zca} (for a review see~\cite{Desjacques:2016bnm}). In this section we give explicit formulas for all relevant one-loop contributions. We will follow the notation of~\cite{Assassi:2014fva}.
%\vskip 4pt

The density contrast of biased tracers, such as the dark matter halos $\delta_h$, is a local function of the underlying dark matter field. The functional dependence is expressed through all possible operators built from the tidal tensor $\partial_i\partial_j\Phi$ (and its derivatives), where $\Phi$ can be either gravitational potential $\Phi_g$ or velocity potential $\Phi_v$. These two potentials are the same at leading order in perturbation theory but starting from second order they differ. To calculate the one-loop power spectrum of biased tracers one has to keep in bias expansion all operators up to third order in perturbation theory\footnote{Notice that \cite{Assassi:2014fva} is using $b_{\delta_2}=\tfrac{b_2}2$ and $b_{\delta_3}=\tfrac{b_3}6$.}
\be
\delta_{h} = b_1 \delta +\frac{b_2}2 \delta^2 +b_{\G_2} \G_2 + \frac{b_3}{6}\delta^3 + b_{\G_3} \G_3 + b_{(\G_2\delta)} \G_2\delta + b_{\Gamma_3} \Gamma_3\;.
\ee
The operators $\G_2$, $\G_3$ and $\Gamma_3$ are defined as
\begin{align}
\G_2(\Phi) & \equiv (\partial_i\partial_j\Phi)^2-(\partial^2\Phi)^2 \;, \\
\G_3(\Phi) & \equiv -\partial_i\partial_j \Phi\, \partial_j\partial_k\Phi\, \partial_k \partial_i \Phi -\frac 12 (\partial^2\Phi)^2 +\frac 32 (\partial_i\partial_j\Phi)^2\partial^2\Phi \;, \\
\Gamma_3(\Phi_g,\Phi_v) & \equiv \G_2(\Phi_g) - \G_2(\Phi_v) \; .
\end{align}
However, only four renormalized operators contribute to the one-loop power spectrum. These are $\delta$, $[\delta^2]$, $[\G_2]$ and $[\Gamma_3]$. The final answer is given in terms of four corresponding renormalized bias parameters and six different momentum integrals \cite{Assassi:2014fva}
\begin{figure}[h]
\includegraphics[width=0.5\textwidth]{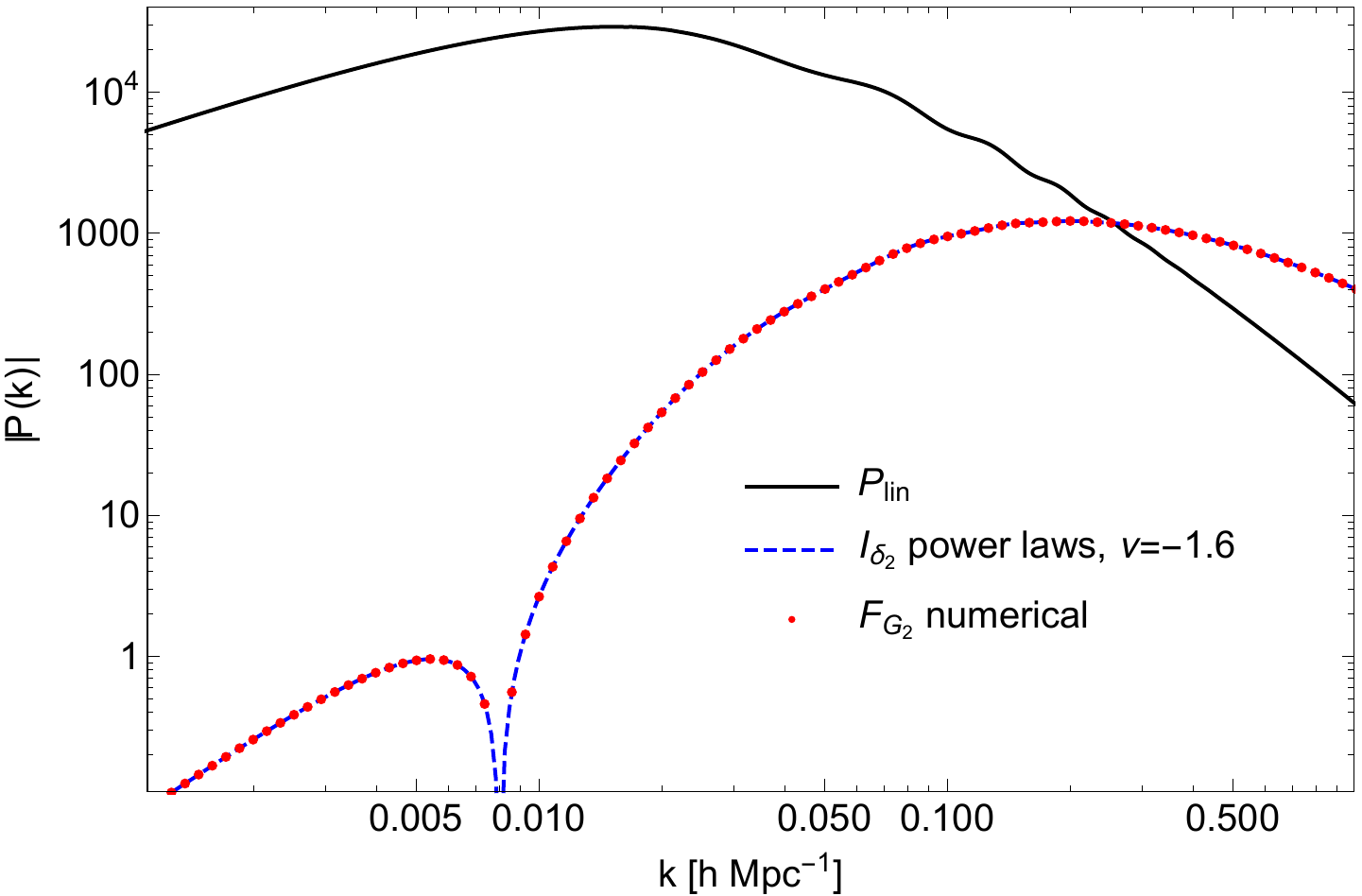}
\includegraphics[width=0.5\textwidth]{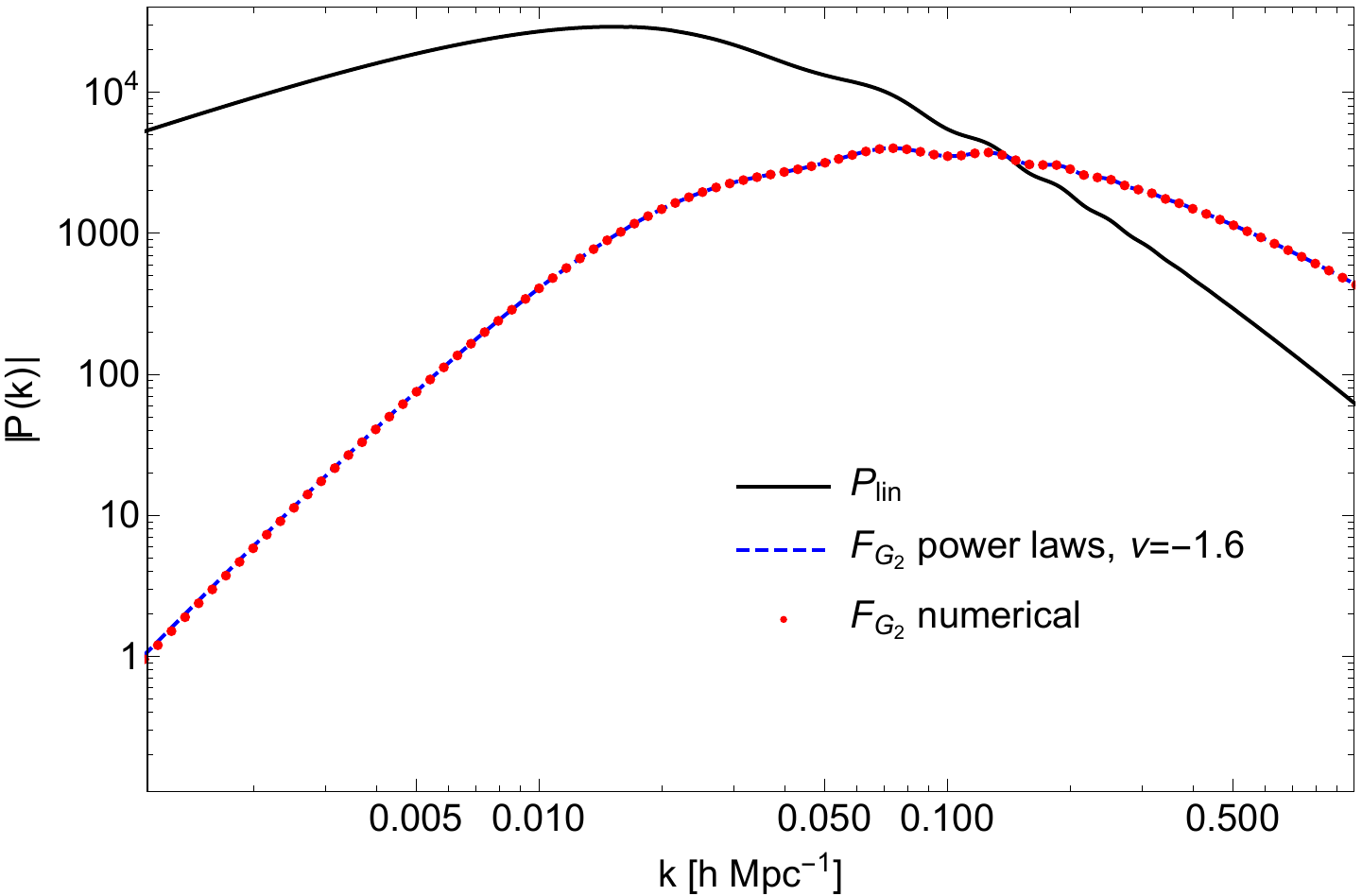}
\includegraphics[width=0.5\textwidth]{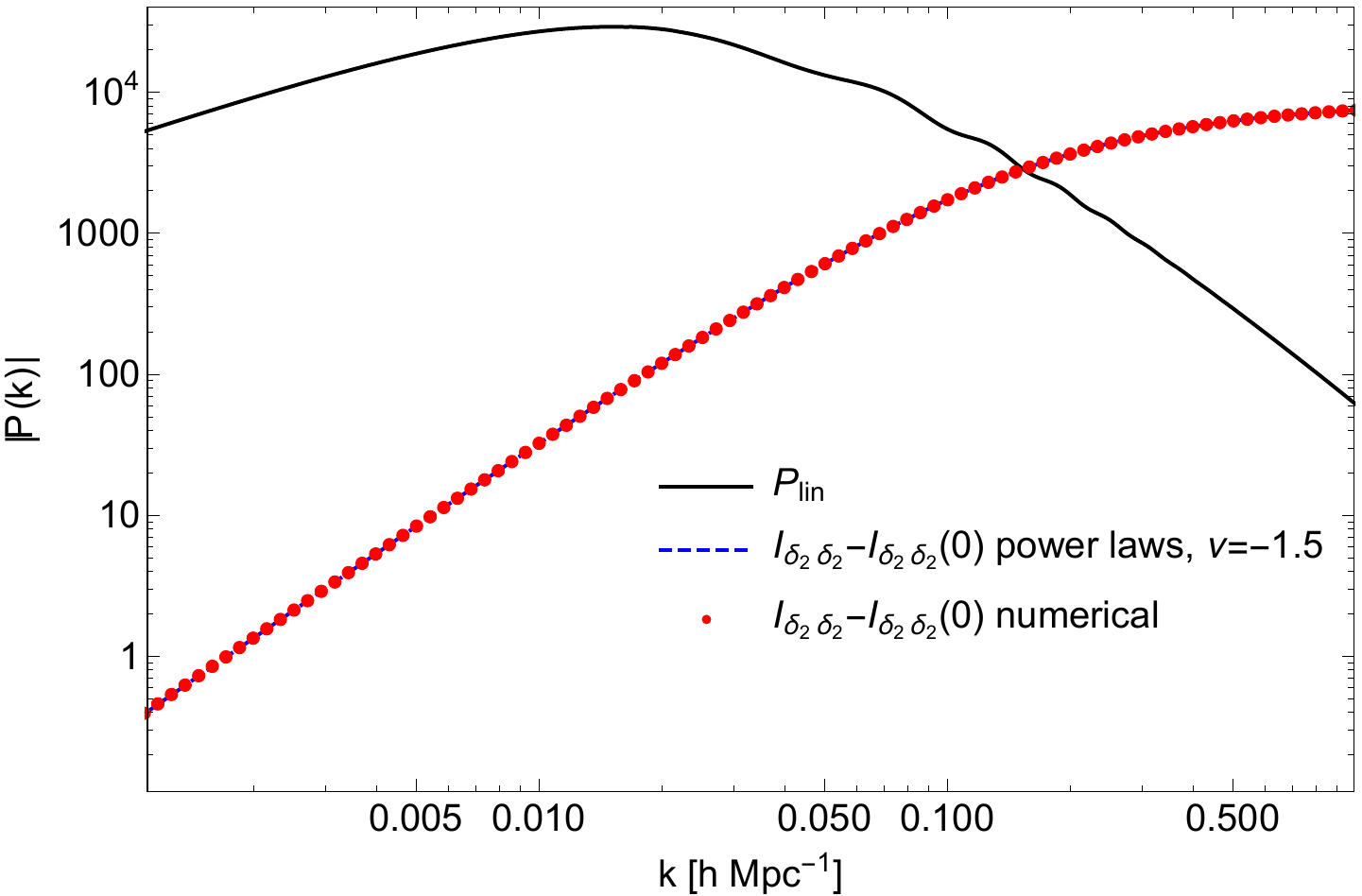}
\includegraphics[width=0.5\textwidth]{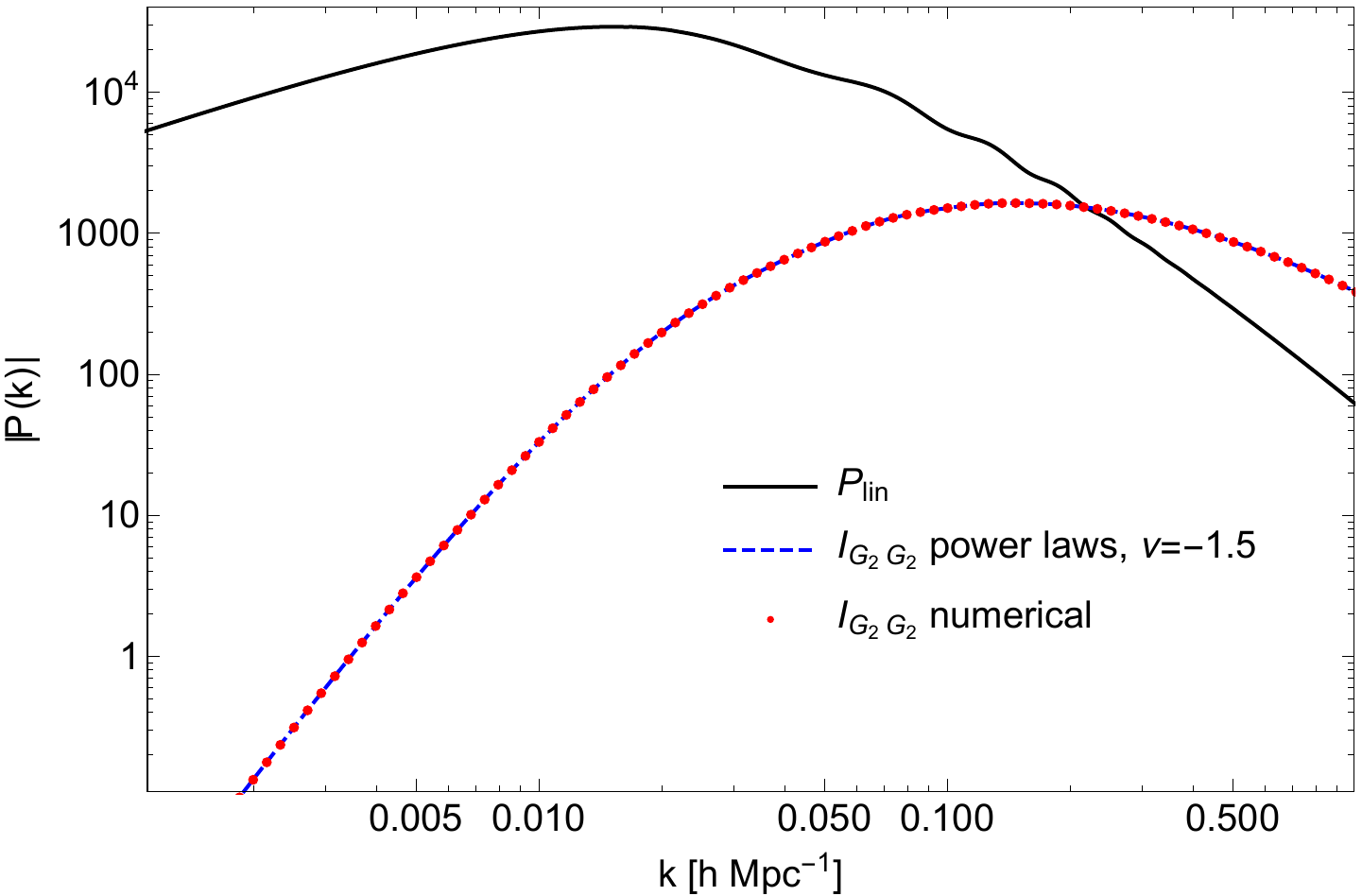}
\caption{Four different contributions to the one-loop power spectrum of biased tracers. All plots are produced using $\nu=-1.6$, $N=150$, $k_{\rm min}=10^{-5}\,h{\rm Mpc}^{-1}$ and $k_{\rm max}=5\,h{\rm Mpc}^{-1}$. For these values of parameters the difference with respect to the usual numerical calculation is less than $0.1\%$ at all scales.}
\label{fig:pbias}
\end{figure}
\begin{align}
P_h(k,\tau) =& b_1^2(P_{\rm lin}(k,\tau)+ P_{\rm 1-loop}(k,\tau))  \nonumber \\
& + b_1 b_2 \,\mathcal I_{\delta^2}(k,\tau) +  2b_1 b_{\G_2} \,\mathcal I_{\G_2}(k,\tau) + \left( 2b_1 b_{\G_2} + \frac 45 b_1 b_{\Gamma_3}\right) \mathcal F_{\G_2}(k,\tau) \nonumber \\
& + \frac 14 b_2^2 \, \mathcal I_{\delta^2\delta^2}(k,\tau) + b_{\G_2}^2 \, \mathcal I_{\G_2\G_2} (k,\tau) + \frac 12 b_2b_{\G_2} \,\mathcal I_{\delta_2\G_2}(k,\tau) \;.
\end{align}
In principle, at this order in perturbation theory one has to add higher derivative operators such as $\partial^2\delta$. However the contribution from this operator is trivial and it does not lead to a loop integral.

The time dependence of all momentum integrals in $P_h(k,\tau)$ is $D(\tau)^4$. The explicit $k$-dependences are 
\begin{align}
\mathcal I_{\delta^2}(k) & = 2\int_{\q} F_2(\q,\k-\q)P_{\rm lin}(q)P_{\rm lin}(|\k-\q|)\, , & \left(-3<\nu<-\tfrac 12\right) \\ 
\mathcal I_{\G_2}(k) & = 2 \int_{\q} \sigma^2(\q,\k-\q)F_2(\q,\k-\q) P_{\rm lin}(q)P_{\rm lin}(|\k-\q|)\, , & \left(-3<\nu<\tfrac 12\right) \\
\mathcal F_{\G_2}(k) & = 4 P_{\rm lin}(k) \int_{\q} \sigma^2(\q,\k-\q)F_2(\k,-\q) P_{\rm lin}(q)\, , & \left(-3<\nu<-1\right) \\
\mathcal I_{\delta^2\delta^2}(k) &= 2 \int_{\q}P_{\rm lin}(q)P_{\rm lin}(|\k-\q|)\, , & \left(-3<\nu<-\tfrac 32\right) \\
\mathcal I_{\G_2\G_2} (k) & = 2 \int_{\q} (\sigma^2(\q,\k-\q))^2 P_{\rm lin}(q)P_{\rm lin}(|\k-\q|)\, , & \left(-3<\nu<\tfrac 12\right) \\
\mathcal I_{\delta_2\G_2}(k) &=  2 \int_{\q} \sigma^2(\q,\k-\q) P_{\rm lin}(q)P_{\rm lin}(|\k-\q|)\, , & \left(-3<\nu<-\tfrac 12 \right)
\end{align}
where $\sigma^2(\k_1,\k_2)=(\k_1\cdot\k_2/k_1k_2)^2-1$. For each term we give a range of power laws for which the integral is convergent. Following the same steps as in the case of the one-loop power spectrum of matter fluctuations, we find that matrices analogous to $M_{22}$ and $M_{13}$ are given by 
\begin{align}
M_{\mathcal I_{\delta^2}}(\nu_1,\nu_2) &= \frac{(3-2\nu_{12})(4-7\nu_{12})}{17\nu_1\nu_2}\;\I(\nu_1,\nu_2)\,, \\
M_{\mathcal I_{\G_2}}(\nu_1,\nu_2) &= -\frac{(3-2\nu_{12})(1-2\nu_{12})(6+7\nu_{12})}{28\nu_1(1+\nu_1)\nu_2(1+\nu_2)}\;\I(\nu_1,\nu_2)\,, \\
M_{\mathcal F_{\G_2}}(\nu_1) &= -\frac{15\tan(\nu_1\pi)}{28\pi(\nu_1+1)\nu_1(\nu_1-1)(\nu_1-2)(\nu_1-3)}\,, \\
M_{\mathcal I_{\delta^2\delta^2}}(\nu_1,\nu_2) &= 2 \I(\nu_1,\nu_2)\,, \\
M_{\mathcal I_{\G_2\G_2}}(\nu_1,\nu_2) &= \frac{(3-2\nu_{12})(1-2\nu_{12})}{\nu_1(1+\nu_1)\nu_2(1+\nu_2)}\;\I(\nu_1,\nu_2)\,, \\
M_{\mathcal I_{\delta^2\G_2}}(\nu_1,\nu_2) &= \frac{3-2\nu_{12}}{\nu_1\nu_2}\;\I(\nu_1,\nu_2)\,.
\end{align}
In Fig.~\ref{fig:pbias} we plot some of the shapes and compare our method with the standard numerical evaluation.  Notice that for $\mathcal I_{\delta^2\delta^2}(k)$ shape we subtract the constant shot-noise part and plot just the difference $\mathcal I_{\delta^2\delta^2}(k)-\mathcal I_{\delta^2\delta^2}(0)$. This difference is convergent even for $\nu>-\tfrac 32$. 
%\vskip 4pt

One important point to make is that the full one-loop power spectrum of biased tracers requires only a single function $\I(\nu_1,\nu_2)$ with a single bias in the range $-3<\nu<-\tfrac 32$. This range can be extended to higher biases by adding the corresponding UV parts of the integrals in the same way as for the matter power spectrum. 
\vskip 10pt

\section{One-loop Bispectrum}
\begin{figure}[h]
\begin{center}
\includegraphics[width=1.0\textwidth]{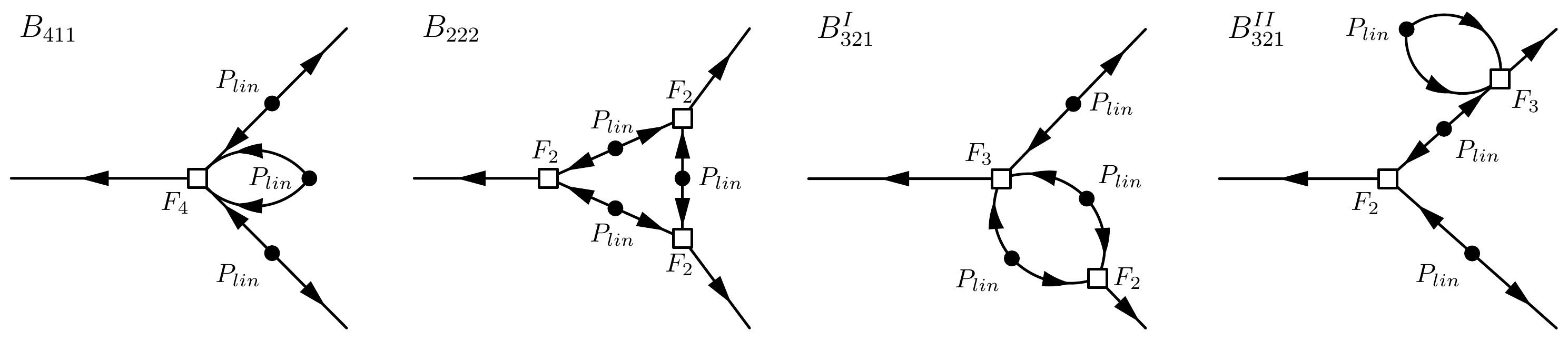}
\end{center}
\caption{Diagrammatic representation of four contributions to the one-loop bispectrum.}
\label{fig:diagrams1LB}
\end{figure}

In perturbation theory there are four different diagrams that contribute to the one-loop bispectrum and their sum can be schematically written like~\cite{Bernardeau:2001qr, Scoccimarro:1996jy, Baldauf:2014qfa, Angulo:2014tfa}
\be
\label{eq:B1loopdef}
B_{\rm 1-loop}(\k_1,\k_2,\k_3,\tau) = D^4(\tau)[B_{222}+B_{321}^I+B_{321}^{II}+B_{411}]\;.
\ee
From translational invariance it follows that $\k_1+\k_2+\k_3 = 0$. The individual terms in square brackets are given by the following integrals
\begin{align}
B_{222} = 8 \int_{\q} F_2 (\q,\k_1-\q) & F_2(\k_1-\q,\k_2+\q) F_2(\k_2+\q,-\q) \nonumber \\
& \times P_{\rm lin}(q) P_{\rm lin}(|\k_1-\q|) P_{\rm lin}(|\k_2+\q|) \;,
\end{align}
\be
B_{321}^I = 6 P_{\rm lin}(k_1) \int_{\q} F_3(\q,\k_2-\q,-\k_1) F_2(\q,\k_2-\q) P_{\rm lin}(q) P_{\rm lin}(|\k_2-\q|) + 5\;{\rm perms} \;,
\ee

\be
B_{321}^{II} = F_2(\k_1,\k_2) P_{\rm lin}(k_1) P_{13}(k_2) + 5\;{\rm perms} \;,
\ee

\be
B_{411} = 12 P_{\rm lin}(k_1) P_{\rm lin}(k_2) \int_{\q} F_4(\q,-\q,-\k_1,-\k_2) P_{\rm lin}(q) + 2\;{\rm cyclic\; perms} \;.  
\ee
The diagrammatic representation of all these contributions is shown in Fig.~\ref{fig:diagrams1LB}.

\begin{figure}[h]
\begin{center}
\includegraphics[width=0.8\textwidth]{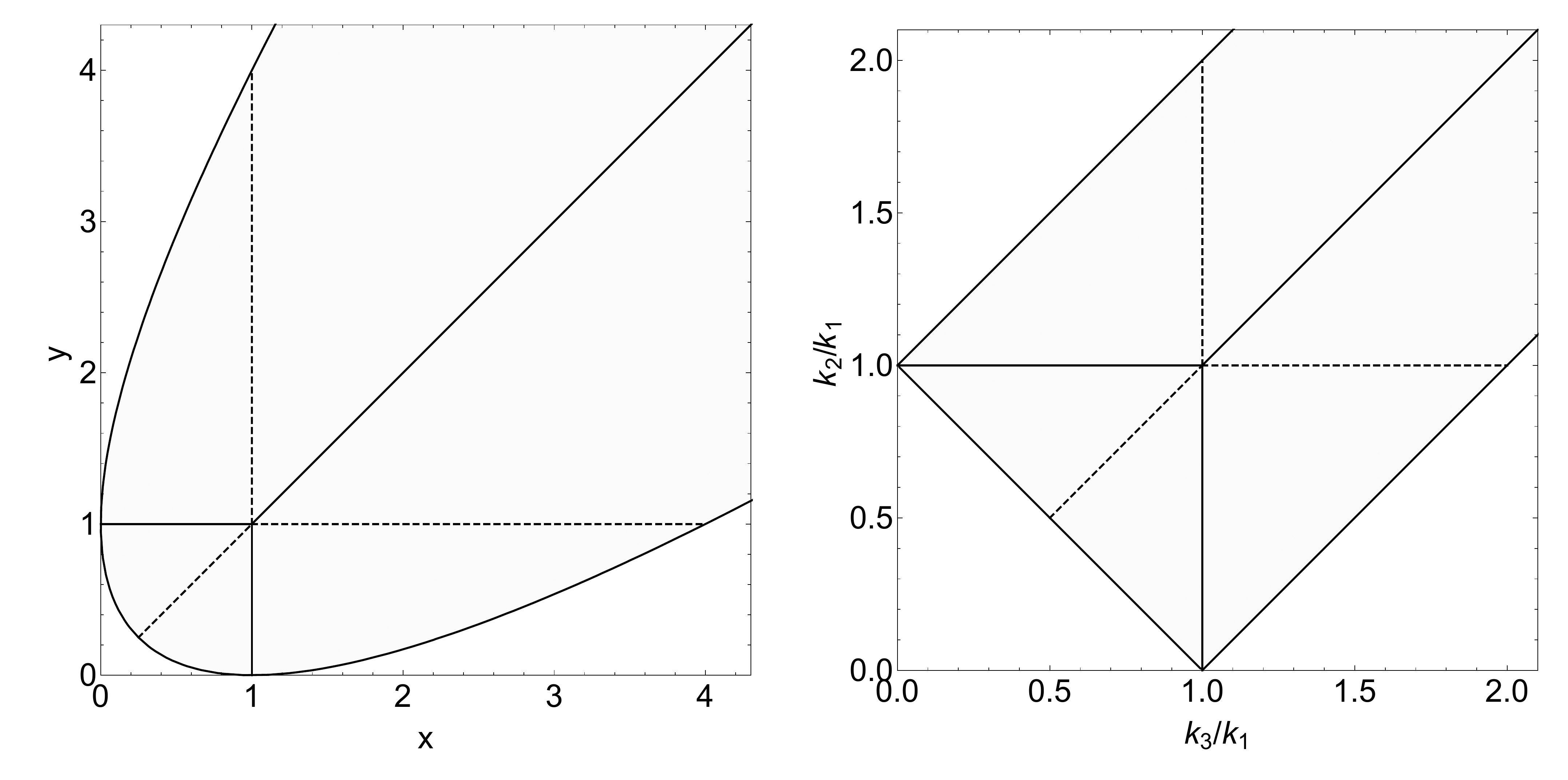}
\end{center}
\caption{{\em Left panel:} Domain of $\J(\nu_1,\nu_2,\nu_3;x,y)$ allowed by the triangle inequality. Six different regions correspond to six permutations of external momenta. All bispectrum configurations can be evaluated focusing on one of these regions. For example, in this paper we choose $x\leq y\leq 1$ which corresponds to $k_3\leq k_2\leq k_1$. Solid lines split the domain in three different parts that will be relevant for the evaluation of the two-loop power spectrum (see Section~\ref{sec:2looppowerevaluation}). {\em Right panel:} The same as left panel in more conventional variables $k_3/k_1$ and $k_2/k_1$.}
\label{fig:domain}
\end{figure}

To evaluate the one-loop bispectrum we can follow the same steps as for the one-loop power spectrum. After expanding the kernels and decomposing the linear power spectrum in power laws, all terms in the sums are proportional to the integral of the following form \cite{Scoccimarro:1996jy}
\be
\label{eq:Jintdef}
\int_{\q} \frac{1}{q^{2\nu_1}|\k_1-\q|^{2\nu_2} |\k_2+\q|^{2\nu_3} } \equiv k_1^{3-2\nu_{123}}\,\J(\nu_1,\nu_2,\nu_3;x,y) \;,
\ee
where $x\equiv k_3^2/k_1^2$ and $y\equiv k_2^2/k_1^2$. Parameters $
\nu_1$, $\nu_2$ and $\nu_3$ have the same form as before (see~\eqref{eq:formnui}). The overall scaling of the integral with momentum is fixed and here we choose to express that scaling in terms of $k_1$. The rest defines a function $\J(\nu_1,\nu_2,\nu_3;x,y)$ which depends only on the ratios $x$ and $y$. Triangle inequality $|k_2-k_3|\leq k_1\leq k_2+k_3$ implies that the physically  allowed region in $(x,y)$ plane is given by inequalities $|\sqrt{x}-\sqrt y|\leq 1$ and $\sqrt x+\sqrt y \geq 1$ and we will focus on evaluating the function in this region (see Fig.~\ref{fig:domain}). Before giving the explicit expression suitable for numerical evaluation we present some important symmetry properties of $\J(\nu_1,\nu_2,\nu_3;x,y)$ which can be derived from its integral representation.

\vskip 10pt

\subsection{Symmetries of $\J(\nu_1,\nu_2,\nu_3;x,y)$ and Recursion Relations}
\label{sec:bispectrumsym}
As in the case of the one-loop power spectrum, the simplest identities follow from shifts of the momentum $\q$. There are two basic translation formulas for function $\J(\nu_1,\nu_2,\nu_3;x,y)$. The first one follows from $\q\to\k_1-\q$ and reads 
\be
\J\left(\nu_1,\nu_2,\nu_3;x,y\right) = \J\left(\nu_2,\nu_1,\nu_3;y,x\right) \;.
\ee
If we do a different shift, $\q \to \q - \k_2$, we get 
\be
\J\left(\nu_1,\nu_2,\nu_3;x,y\right)=x^{3/2-\nu_{123}} \J\left(\nu_3,\nu_2,\nu_1;\tfrac{1}{x},\tfrac{y}{x}\right) \;.
\ee
These two formulas are sufficient to generate identities involving all six permutations of parameters $\nu_1$, $\nu_2$ and $\nu_3$. These are
\be
\label{eq:Jpermutationsym}
\begin{split}
\J\left(\nu_1,\nu_2,\nu_3; x,y\right)=&\J\left(\nu_2,\nu_1,\nu_3; y,x\right)\\
=&x^{3/2-\nu_{123}} \J\left(\nu_3,\nu_2,\nu_1; \tfrac{1}{x},\tfrac{y}{x}\right)=x^{3/2-\nu_{123}} \J\left(\nu_2,\nu_3,\nu_1; \tfrac{y}{x},\tfrac{1}{x}\right)\\
=&y^{3/2-\nu_{123}} \J\left(\nu_3,\nu_1,\nu_2; \tfrac{1}{y},\tfrac{x}{y}\right)=y^{3/2-\nu_{123}} \J\left(\nu_1,\nu_3,\nu_2; \tfrac{x}{y},\tfrac{1}{y}\right) \;.
\end{split}
\ee 
An intuitive way to understand these symmetries is to realize that they map $\k_1$, $\k_2$ and $\k_3$ into each other, preserving the shape of the triangle. The six equations then correspond to nothing other but six possible permutations of three external momenta. From another point of view, for evaluation of the bispectrum we can always choose a small ``corner'' in the $(x,y)$ plane (see Fig.~\ref{fig:domain}). The choice that we make in this paper is $x\leq y\leq 1$ which corresponds to the following ordering of momenta: $k_3\leq k_2\leq k_1$. 
%\vskip 4pt

Let us now derive the inversion formula for $\J(\nu_1,\nu_2,\nu_3;x,y)$. We can start from 
\be
\J(\nu_1,\nu_2,\nu_3;x,y) = \int_{\q} \frac{1}{q^{2\nu_1}|\hat\k_1-\q|^{2\nu_2} |\sqrt y\hat \k_2+\q|^{2\nu_3} } \;,
\ee
where $\hat\k_1$ and $\hat\k_2$ are unit vectors. Under inversion $\q\to\q/q^2$, apart from transformations described in~\eqref{eq:Iinvtransf}, we also get 
\be
|\sqrt y\hat \k_2+\q|^{2} \;\; \to \;\; \frac{y}{q^2} |\hat\k_2/\sqrt y + \q |^2 \;.
\ee
The whole integral then changes to 
\be
\J(\nu_1,\nu_2,\nu_3;x,y) = \int_{\q} \frac{y^{-\nu_3}}{q^{2(3-\nu_{123})}|\hat\k_1-\q|^{2\nu_2} |\hat \k_2/\sqrt y+\q|^{2\nu_3} } \;.
\ee
It is easy to read off the inversion formula from this expression. One only has to keep in mind that, due to $\sqrt y$ appearing in the denominator in the last term, the arguments of the function change to $\tfrac xy$ and $\tfrac 1y$. We finally get
\be
\label{eq:Jinversionformula}
\J(\nu_1,\nu_2,\nu_3;x,y) =  y^{-\nu_3} \J\left(3-\nu_{123},\nu_2,\nu_3;\tfrac xy,\tfrac 1y \right) \;. 
\ee
Combining this inversion formula with translation formulas~\eqref{eq:Jpermutationsym}, we get a set of three identities for functions evaluated at the {\em same} point $(x,y)$
\be
\label{eq:Jpermutationsym1}
\begin{split}
\J(\nu_1,\nu_2,\nu_3;x,y)=&\, x^{3/2-\nu_{23}}\, \J(\nu_3,3-\nu_{123},\nu_1;x,y)\\
=& \, x^{3/2-\nu_{23}} \, y^{3/2-\nu_{13}}\, \J(\nu_2,\nu_1,3-\nu_{123};x,y) \\
=&\, y^{3/2-\nu_{13}} \, \J(3-\nu_{123},\nu_3,\nu_2;x,y) \;.
\end{split}
\ee 
%\vskip 4pt 

When $\nu_{123}=3$ the previous equation implies 
\be
\label{eq:star-triangle}
\J(\nu_1,\nu_2,3-\nu_{12}) = x^{3/2-\nu_{23}} \, y^{3/2-\nu_{13}}\,\I(\nu_1,\nu_2) \;.
\ee
This expression is sometimes referred to as the {\em star-triangle} duality. The condition $\nu_{123}=3$ does not correspond to a generic situation but it can be used to derive another generic formula for $\J(\nu_1,\nu_2,\nu_3;x,y)$. The idea is to ``split'' one of the parameters using the one-loop integral, such that $\nu_{123}=3$ is satisfied. For example, we can choose to write 
\be
\frac1{|\k_2+\q|^{2\nu_3}} = \frac1{\I(\nu_{123}-\tfrac 32, 3-\nu_{12})} \int_\s \frac1{s^{2\nu_{123}-3}|\k_2+\q-\s|^{6-2\nu_{12}}} \;.
\ee
Now one can start with the integral representation of $\J(\nu_1,\nu_2,\nu_3;x,y)$ and use the previous formula. Notice that we have chosen parameters such that the integration in $\q$ can be then easily done using~\eqref{eq:star-triangle}. The remaining integral in $\s$ has again the form of the one-loop bispectrum. Following these steps one derives the star-triangle formula 
\be
\label{eq:Jstartriangle}
\J(\nu_1,\nu_2,\nu_3;x,y) = \tfrac{\Gamma(\tilde\nu_1)}{\Gamma(\nu_1)} \tfrac{\Gamma(\tilde\nu_2)}{\Gamma(\nu_2)} \tfrac{\Gamma(\tilde\nu_3)}{\Gamma(\nu_3)} \tfrac{\Gamma(3-\tilde\nu_{123})}{\Gamma(3-\nu_{123})} \, \J(\tilde\nu_2,\tilde\nu_1,3-\tilde\nu_{123} ;x,y) \;,
\ee
where $\tilde\nu_i = \tfrac32-\nu_i$. Finally, using the method described after eq.~\eqref{eq:duality}, it is possible to derive the following reflection formula 
\be
\label{eq:Jreflection}
\J(\nu_1,\nu_2,\nu_3;x,y) =  \tfrac{\Gamma(\tilde\nu_1)}{\Gamma(\nu_1)} \tfrac{\Gamma(\tilde\nu_2)}{\Gamma(\nu_2)} \tfrac{\Gamma(\tilde\nu_3)}{\Gamma(\nu_3)} \tfrac{\Gamma(3-\tilde\nu_{123})}{\Gamma(3-\nu_{123})} x^{3/2-\nu_{23}} y^{3/2-\nu_{13}} \J(\tilde\nu_1,\tilde\nu_2, \tilde\nu_3 ; x,y) \;.
\ee 
This is not an independent relation because it follows from eq.~\eqref{eq:Jpermutationsym1} and eq.~\eqref{eq:Jstartriangle}. 
%\vskip 4pt

For arbitrary choice of $\nu$ eq.~\eqref{eq:Jpermutationsym1} and eq.~\eqref{eq:Jreflection} are not very useful, because they relate two functions with two {\em different} biases. However, there are some special choices of $\nu$ for which this is not the case. Let us remember that, up to an integer, the structure of parameters is 
\be
\nu_i = -\frac \nu2 - i \frac{\eta_{m_i}}2 \;,
\ee
which implies 
\be
\tilde \nu_i = \frac 32 + \frac \nu2 + i \frac{\eta_{m_i}}2 \;.
\ee
Up to an integer, the transformation $\nu\to\tilde\nu$ does not change the bias if $\nu$ is a odd integer multiple of $-\tfrac 12$. For example, let us imagine that $\nu=-\tfrac32$. In this case
\be
\nu_i = \frac 34 - i \frac{\eta_{m_i}}2  \qquad \Rightarrow \qquad \tilde\nu_i = \frac 34 + i \frac{\eta_{m_i}}2  \;.
\ee
In other words, for $\nu=-\tfrac32$, $\tilde \nu_i$ is just a complex conjugate of $\nu_i$. For example, assuming $\nu=-\tfrac32$, the reflection formula becomes
\begin{align}
\J(\nu_1+n_1, & \nu_2+n_2,\nu_3+n_3 ;x,y) =  \tfrac{\Gamma(\tilde\nu_1-n_1)}{\Gamma(\nu_1+n_1)} \tfrac{\Gamma(\tilde\nu_2-n_2)}{\Gamma(\nu_2+n_2)} \tfrac{\Gamma(\tilde\nu_3-n_3)}{\Gamma(\nu_3+n_3)} \tfrac{\Gamma(3-\tilde\nu_{123}+n_{123})}{\Gamma(3-\nu_{123}-n_{123})} \nonumber \\
& \times x^{3/2-\nu_{23}-n_{23}} y^{3/2-\nu_{13}-n_{13}} \J^*(\nu_1-n_1,\nu_2-n_2, \nu_3-n_3 ; x,y) \;,
\end{align}
where $n_i$ are integers coming from the expansion of the kernels. This equation provides a simple relation between two functions in the same point and with the same parameters, but with the opposite sign of integer part of $\nu_i$. For example, such pairs of functions do exist in the expansion of kernels in $B_{222}$ and each of them can be calculated for the price of a single evaluation. Similar identity can be derived for $\nu=-\tfrac 12$ and the same conclusions apply to a set of formulas~\eqref{eq:Jpermutationsym1}. 
%\vskip 4pt 

Let us conclude showing that the function $\J(\nu_1,\nu_2,\nu_3;x,y)$ satisfies a set of recursion relations~\cite{Davydychev:1992xr}. As in the case of the one-loop power spectrum we can start from the following identity
\be
\label{eq:totalderiv}
\int_{\q} \frac{\partial}{\partial q_i} \left( \frac{q_i}{q^{2\nu_1}|\k_1-\q|^{2\nu_2} |\k_2+\q|^{2\nu_3}} \right) = 0 \;.
\ee
After expanding the derivative in the integrand, the previous equation can be rewritten as 
\begin{align}
\label{eq:Jrecursion1}
& \nu_2 \, \J(\nu_1,\nu_2+1,\nu_3) + \nu_3y\,\J(\nu_1,\nu_2,\nu_3+1) = \nonumber \\
& \quad (\nu_1+\nu_{123} -3) \J(\nu_1,\nu_2,\nu_3) + \nu_2 \, \J(\nu_1-1,\nu_2+1,\nu_3) +\nu_3 \, \J(\nu_1-1,\nu_2,\nu_3+1)\;,
\end{align}
where we suppressed the $(x,y)$ argument in all functions to avoid clutter. There are two other similar expressions that can be derived replacing $q_i$ in the numerator of the integral in \eqref{eq:totalderiv} with $(\k_1-\q)_i$ or $(\k_2+\q)_i$. These are
\begin{align}
& \nu_1 \, \J(\nu_1+1,\nu_2,\nu_3) + \nu_3x\,\J(\nu_1,\nu_2,\nu_3+1) = \nonumber \\
& \quad (\nu_2+\nu_{123} -3) \J(\nu_1,\nu_2,\nu_3) + \nu_1 \, \J(\nu_1+1,\nu_2-1,\nu_3) +\nu_3 \, \J(\nu_1,\nu_2-1,\nu_3+1)\;, \\
& \nu_1y \, \J(\nu_1+1,\nu_2,\nu_3) + \nu_2 x\,\J(\nu_1,\nu_2+1,\nu_3) = \nonumber \\
& \quad (\nu_3+\nu_{123} -3) \J(\nu_1,\nu_2,\nu_3) + \nu_1 \, \J(\nu_1+1,\nu_2,\nu_3-1) +\nu_2 \, \J(\nu_1,\nu_2+1,\nu_3-1)\;.
\end{align}
Notice that we wrote these equations such that the sum of the arguments in each function on the l.h.s.~is $\nu_{123}+1$ and  the sum of the arguments in each function on the r.h.s.~is $\nu_{123}$. This splitting suggests the following interpretation of the recursion relations.
They can be thought of as a system of three linear equations where three unknown functions are those in which one of the parameters is increased by 1. If we denote the r.h.s.~of previous equations with $A_1$, $A_2$ and $A_3$
\begin{align}
A_1\equiv\, & (\nu_1+\nu_{123} -3) \J(\nu_1,\nu_2,\nu_3) + \nu_2 \, \J(\nu_1-1,\nu_2+1,\nu_3) +\nu_3 \, \J(\nu_1-1,\nu_2,\nu_3+1) \;, \\
A_2\equiv\, & (\nu_2+\nu_{123} -3) \J(\nu_1,\nu_2,\nu_3) + \nu_1 \, \J(\nu_1+1,\nu_2-1,\nu_3) +\nu_3 \, \J(\nu_1,\nu_2-1,\nu_3+1)\;,\\
A_3\equiv\, & (\nu_3+\nu_{123} -3) \J(\nu_1,\nu_2,\nu_3) + \nu_1 \, \J(\nu_1+1,\nu_2,\nu_3-1) +\nu_2 \, \J(\nu_1,\nu_2+1,\nu_3-1)\;,
\end{align}
then the solution of the system is given by~\cite{Davydychev:1992xr}
\begin{align}
\label{eq:Jrecursion}
\begin{split}
\J(\nu_1+1,\nu_2,\nu_3) &= \frac{1}{2\nu_1y} \big(-A_1 x + A_2 y + A_3 \big) \;, \\
\J(\nu_1,\nu_2+1,\nu_3) &= \frac{1}{2\nu_2 x} \big(A_1 x - A_2 y + A_3 \big) \;, \\
\J(\nu_1,\nu_2,\nu_3+1) &= \frac{1}{2\nu_3 x y} \big(A_1 x + A_2 y - A_3 \big) \;.
\end{split}
\end{align}
In other words, seven different functions whose parameters live on the plane $\nu_{123}={\rm const.}$ determine three extra integrals on the plane $\nu_{123}+1={\rm const}$. These identities are very useful. For example, in the expansion of kernels in $B_{222}$ they reduce the number of independent terms by roughly a factor of 2 (from 72 to 38). They are also very important for simplifying the two-loop calculation as we are going to see in the following sections. 
\vskip 10pt

%
%\begin{align}
%2\nu_1y\, J(\nu_1+1,\nu_2,\nu_3) =& \big( -(\nu_1+\nu_{123} -3)x + (\nu_2 + \nu_{123} -3)y + (\nu_3+\nu_{123} -3)\big) J(\nu_1,\nu_2,\nu_3)  \nonumber \\
%& - \nu_2 x \, J(\nu_1-1,\nu_2+1,\nu_3) - \nu_3x  \, J(\nu_1-1,\nu_2,\nu_3+1)
%\nonumber \\
%& + \nu_1y\, J(\nu_1+1,\nu_2-1,\nu_3) +\nu_3 y \, J(\nu_1,\nu_2-1,\nu_3+1)
%\nonumber \\
%& + \nu_1 J(\nu_1+1,\nu_2,\nu_3-1) + \nu_2 J(\nu_1,\nu_2+1,\nu_3-1) \;, 
%\end{align}
%\begin{align}
%2\nu_2 x \, J(\nu_1,\nu_2+1,\nu_3) =& \big( (\nu_1+\nu_{123} -3)x - (\nu_2 + \nu_{123} -3)y + (\nu_3+\nu_{123} -3)\big) J(\nu_1,\nu_2,\nu_3)  \nonumber \\
%& +\nu_2 x \, J(\nu_1-1,\nu_2+1,\nu_3) +\nu_3x  \, J(\nu_1-1,\nu_2,\nu_3+1)
%\nonumber \\
%& - \nu_1y\, J(\nu_1+1,\nu_2-1,\nu_3) -\nu_3 y \, J(\nu_1,\nu_2-1,\nu_3+1)
%\nonumber \\
%& + \nu_1 J(\nu_1+1,\nu_2,\nu_3-1) + \nu_2 J(\nu_1,\nu_2+1,\nu_3-1) \;, 
%\end{align}
%\begin{align}
%2\nu_3\, x y\, J(\nu_1,\nu_2,\nu_3+1) =& \big( (\nu_1+\nu_{123} -3)x + (\nu_2 + \nu_{123} -3)y -(\nu_3+\nu_{123} -3)\big) J(\nu_1,\nu_2,\nu_3)  \nonumber \\
%& +\nu_2 x \, J(\nu_1-1,\nu_2+1,\nu_3) +\nu_3x  \, J(\nu_1-1,\nu_2,\nu_3+1)
%\nonumber \\
%& + \nu_1y\, J(\nu_1+1,\nu_2-1,\nu_3) +\nu_3 y \, J(\nu_1,\nu_2-1,\nu_3+1)
%\nonumber \\
%& - \nu_1 J(\nu_1+1,\nu_2,\nu_3-1) - \nu_2 J(\nu_1,\nu_2+1,\nu_3-1) \;, 
%\end{align}

\subsection{Evaluation of $\J(\nu_1,\nu_2,\nu_3;x,y)$}
\label{sec:bispectrumevaluation}
After making these general remarks based on the integral representation, let us turn to the explicit expression for $\J(\nu_1,\nu_2,\nu_3;x,y)$. Unlike $\I(\nu_1,\nu_2)$, this function cannot be simply expressed in a closed form in terms of gamma functions. Starting from~\eqref{eq:Jintdef} and using Feynman parameters we get (see Appendix~\ref{app:J})
\begin{align}
\J(\nu_1,\nu_2, \nu_3 ;x,y) & = \frac{1}{8\pi^{3/2}} \frac{\Gamma\left(\nu_{123}-\frac32 \right)}{\Gamma(\nu_1)\Gamma(\nu_2)\Gamma(\nu_3)} \nonumber \\
& \quad \times \int_{0}^1 du \int_0^1 dv \frac{u^{\nu_1-1}(1-u)^{\nu_2-1}v^{1/2-\nu_3}(1-v)^{\nu_3-1}}{\left( u v (1-u) + u(1-v) y + (1-u) (1-v) x \right)^{\nu_{123}-3/2}} \;.
\end{align}
The expression on the r.h.s.~belongs to the class of hypergeometric functions of two variables. In particular, $\J(\nu_1,\nu_2, \nu_3 ;x,y)$ can be written as a linear combination of Appell $F_4$ functions~\cite{Davydychev:1992xr}. These special functions can be evaluated using their series representations. The region of convergence is given by $\sqrt{x}+\sqrt{y}< 1$, which unfortunately covers only the unphysical part of the $(x,y)$ plane. As usual, this kind of problems can be circumvented by performing the analytic continuation. This can be done in several ways, depending on the region of parameter space that one wants to cover~\cite{Exton:1994ab}. Although all results are formally equivalent and  can be related to each other, distinct expressions can be very different from the point of view of practical calculation. A series representation of $\J(\nu_1,\nu_2, \nu_3 ;x,y)$, optimized for numerical evaluation of the bispectrum, is given by the following formula 
\begin{align}
\label{eq:Jseries}
\J(\nu_1,\nu_2, & \nu_3 ;x,y) = \frac{\sec(\pi\nu_{23})}{8\sqrt \pi \Gamma(\nu_1) \Gamma(\nu_2) \Gamma(\nu_3) \Gamma(3-\nu_{123})}\nonumber \\
&\left[ x^{3/2-\nu_{23}} \sum_{n=0}^\infty  a_{n}(\nu_1,\nu_2,\nu_3)\cdot\; x^{n}\;_2F_1\left(\nu_1+n, \tfrac 32-\nu_2+n, 3-\nu_{23}+2n, 1-y \right) \right. \nonumber \\
& \left. -y^{3/2-\nu_{13}} \sum_{n=0}^\infty a_{n}(\tilde\nu_1,\tilde\nu_2,\tilde\nu_3)\cdot\; x^{n}\;_2F_1\left(\tilde\nu_1+n, \tfrac 32-\tilde\nu_2+n, 3-\tilde\nu_{23}+2n, 1-y \right) \right] \;,
\end{align}
where
\be
a_{n}(\nu_1,\nu_2,\nu_3)=\frac{\Gamma \left(\nu_1+ n\right)\Gamma \left(3 -\nu_{123} + n\right)}{\Gamma \left(\frac 52-\nu_{23}+ n\right)n!} \frac{\Gamma(\frac 32-\nu_3+n)\Gamma(\frac 32-\nu_2+n)}{\Gamma(3-\nu_{23}+2n)} \;.
\ee
The functions ${}_2F_1(\ldots,1-y)$ that appear in the result are standard Gauss hypergeometric functions. We review their definition and some important properties in Appendix~\ref{app:Hyper}. The derivation of eq.~\eqref{eq:Jseries} is given in Appendix~\ref{app:J}.  
%\vskip 4pt

Let us make some comments about expression \eqref{eq:Jseries}. The first thing to notice is that the series is always convergent if we restrict ourselves to the region $x\leq y\leq 1$. The minimal allowed value of $1-y$ in the given region is $\tfrac34$, which corresponds to folded triangles. The smaller $1-y$ is, the easier it gets to calculate the hypergeometric functions using their power series representation. For higher values of $y$ and smaller values of $x$ the convergence is very fast. In the limits of $x\to 0$ and $y\to 1$,  corresponding  to squeezed triangles, only a few terms need be kept in the sum. The slowest convergence is for high values of $x$. The limiting case is $x=1$ and $y=1$, which corresponds to equilateral triangles. Even in this case only a relatively modest number of terms, $\sim \mathcal O(50)$, need be kept in the sum to reach satisfactory precision.
%\vskip 4pt

Another important point to keep in mind is that dependences on $x$ and $y$ are explicitly separated in our formula. Furthermore, the $x$-dependence is trivial. This means that in practice, for a given $y$, calculation for any $k_1$ and $x$ can be done evaluating the hypergeometric functions only once. This can speed up any full bispectrum calculation significantly. 
%\vskip 4pt 

There are additional optimizations which can exploit many well-known properties of hypergeometric functions. One such property is that ${}_2F_1(a+n_1,b+n_2,c+n_3,z)$ for any set of integers $(n_1,n_2,n_3)$ can be always written as a linear combination of just two hypergeometric functions such as, for example, ${}_2F_1(a,b,c,z)$ and ${}_2F_1(a+1,b,c,z)$. Using this property one can prove the following recursion relation\footnote{Notice that in equation $(10)$ of reference~\cite{Morita:1996ab} there is a typo. The sign between the two terms in square brackets should be $+$ instead of $-$.} \cite{Morita:1996ab}
\be
f_{n-1} = f_n +\tfrac{c(1-a-b-2n)+2ab-2n(n-1)}{(c+2n-2)(c+2n)} z \, f_n - \tfrac{(a+n)(b+n)(c-a+n)(c-b+n)}{(c+2n)^2((c+2n)^2-1)}z^2 f_{n+1} \;,
\ee  
where $f_n={}_2F_1(a+n,b+n,c+2n,z)$. This equations gives a way to recursively calculate all hypergeometric functions in the power series of~\eqref{eq:Jseries}. In practice one should exercise some caution, as certain values of arguments in this recursion relation can be numerically unstable. 
%\vskip 4pt

In some special cases eq.~\eqref{eq:Jseries} further simplifies. For example, for the case of isosceles triangles, $y=1$ and all hypergeometric functions are equal to one. As expected from the symmetry properties in~\eqref{eq:Jpermutationsym}, the result in this case becomes symmetric in $\nu_2$ and $\nu_3$.
%\vskip 4pt

Finally, let us point out that if one of the parameters is a negative integer or zero, the sum in~\eqref{eq:Jseries} truncates. To see this explicitly, consider  $\nu_1=-N$, with $N\geq0$ a non-negative integer. In the limit $\nu_1\to-N$, $1/\Gamma(\nu_1)$ in the normalization goes to zero. If there were no terms in the sum which diverge in the same limit, the result would be zero. The hypergeometric functions are always regular. Therefore, we have to look at the coefficients. By inspection we see that all $a_n(\tilde\nu_1,\tilde\nu_2,\tilde\nu_3)$ coefficients are regular as well. The only divergence comes from $\Gamma(\nu_1+n)$ in the coefficient $a_n(\nu_1,\nu_2,\nu_3)$, for $n\leq N$. Given that 
\be
\frac{\Gamma(-N+n)}{\Gamma(-N)} = (-1)^n \frac{\Gamma(N+1)}{\Gamma(N-n+1)} \;, 
\ee
we can rewrite the final answer in the following way
\begin{align}
\label{eq:Jtruncates}
\J(-N,\nu_2, & \nu_3 ;x,y) = (-1)^{N+1} \frac{\sqrt \pi \sec(\pi\nu_{23})\sec(\pi\nu_3)}{8 \Gamma(\nu_2) \Gamma(\nu_3) \Gamma(3+N-\nu_{23})} \nonumber \\
& \sum_{n=0}^N \sum_{m=0}^{N-n} \frac{N!\;(-1)^{m+n} }{(N-m-n)!} \frac{\Gamma(\tfrac 32-\nu_2+n+m)}{\Gamma \left(\frac 52-\nu_{23}+ n\right)\Gamma(\nu_{3} - N -\tfrac 12+m)} \frac{x^{3/2-\nu_{23}+n}}{n!} \frac{y^m}{m!} \;. 
\end{align}
In conclusion, when one of the arguments of $\J(\nu_1,\nu_2,\nu_3 ;x,y)$ is a negative integer or zero, the function becomes a simple polynomial in $x$ and $y$ of degree $N$. If two of the arguments are negative integers or zero, then the function vanishes.
\vskip 10pt

\subsection{Numerical Evaluation of the One-loop Bispectrum}
\label{sec:bispectrumnumerical}
Let us now turn to the numerical evaluation of the bispectrum. We will consider each term in~\eqref{eq:B1loopdef} separately. 
\begin{figure}[h]
\begin{center}
\includegraphics[width=0.7\textwidth]{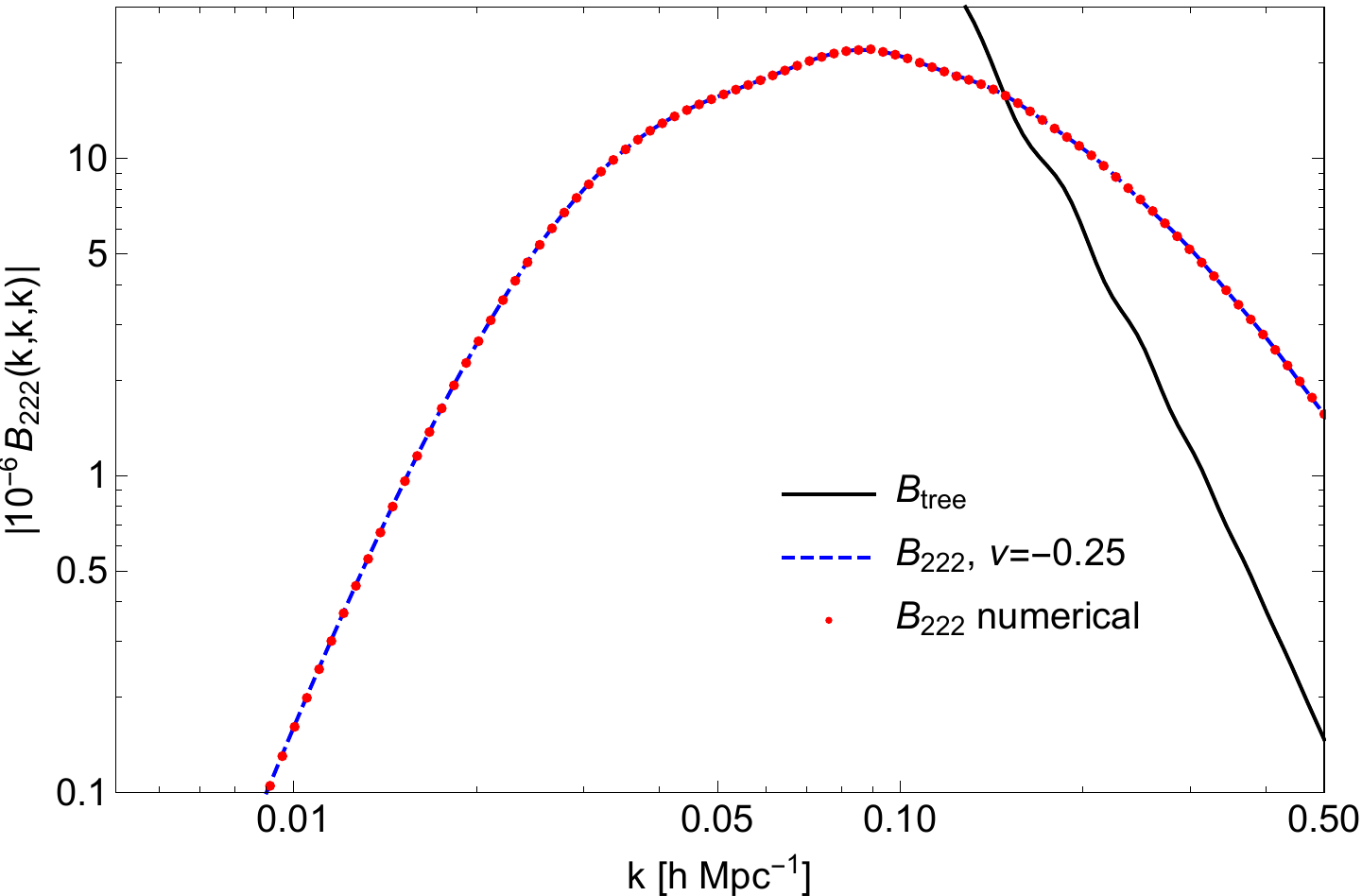}
\end{center}
\caption{Equilateral $B_{222}$ diagram as a function of $k$. The calculation is done with bias $\nu=-0.25$ and $N=50$ sampling points for the power spectrum. This is enough to reach the sub-percent precision.}
\label{fig:b222}
\end{figure}
%\vskip 4pt 

\noindent
{\em $B_{222}$ term.---}We begin with the $B_{222}$ contribution. We will fist find the range of biases for which the integral is convergent. Two of three $F_2$ kernels, have the same structure and asymptotic behavior as in the $P_{22}$ diagram. The third kernel tends to $\mathcal O(1)$ constant in the $\q\to 0$ limit. In the UV limit
\be
F_2(\k_1-\q,\k_2+\q) \to \frac{k^2}{q^2}\;, \qquad q\to\infty \;.
\ee
Combining all these limits it follows that the integral in $B_{222}$ diagram is convergent for power laws in the range $-1<\nu<1$. Therefore, choosing a bias close to zero, we expect our method to reproduce the results of the usual numerical integration. After expanding the kernels and linear power spectra in power laws, we can write the result as a matrix multiplication
\be
\label{eq:B222powers}
\bar B_{222}(k_1,k_2,k_3) = k_1^{3} \sum_{m_i} M_{222}(\nu_1,\nu_2,\nu_3;x,y) \cdot c_{m_1}k_1^{-2\nu_1}\cdot c_{m_2}k_1^{-2\nu_2}\cdot c_{m_3}k_1^{-2\nu_3} \;,
\ee
where the matrix $M_{222}(\nu_1,\nu_2,\nu_3;x,y)$ is given by
\be
M_{222}(\nu_1,\nu_2,\nu_3;x,y) = 8\sum_{n_i} f_{222}(n_1,n_2,n_3;x,y)\,\J(\nu_1-n_1,\nu_2-n_2,\nu_3-n_3;x,y) \;.
\ee
As before, $c_{m_i}$ are the coefficients of the expansion~\eqref{eq:fftlogcoeff} and $n_1$, $n_2$ and $n_3$ are integer powers of $q^2$, $|\k_1-\q|^2$ and $|\k_2+\q|^2$ in the expansion of the three kernels in the integrand. The rational coefficients in this expansion are labeled by $f_{222}(n_1,n_2,n_3;x,y)$. We give the explicit expression for $M_{222}$ in Appendix~\ref{app:Tables}. 
%\vskip 4pt 

Let us make a couple of comments about the formulas above. As in the case of the one-loop power spectrum, the evaluation of the bispectrum boils down to a simple matrix multiplication. The matrix $M_{222}$ is cosmology independent, so it has to be calculated only once. Notice that $M_{222}$ depends only on the shape of the triangle formed by the three external momenta, and not on its absolute size. It is clear from eq.~\eqref{eq:B222powers} that for fixed $x$ and $y$ one can calculate all triangles with different $k_1$ using the same $M_{222}$. The size of this matrix is $N^3$, where $N$ is the number of sampling points of the linear power spectrum. In practice, in order to reach the sub-percent precision, it is enough to use $\mathcal O(50)$ points. We plot the $B_{222}$ diagram in Fig.~\ref{fig:b222} and as expected it is in a very good agreement with the usual numerical evaluation. 
\begin{figure}[h]
\begin{center}
\includegraphics[width=0.7\textwidth]{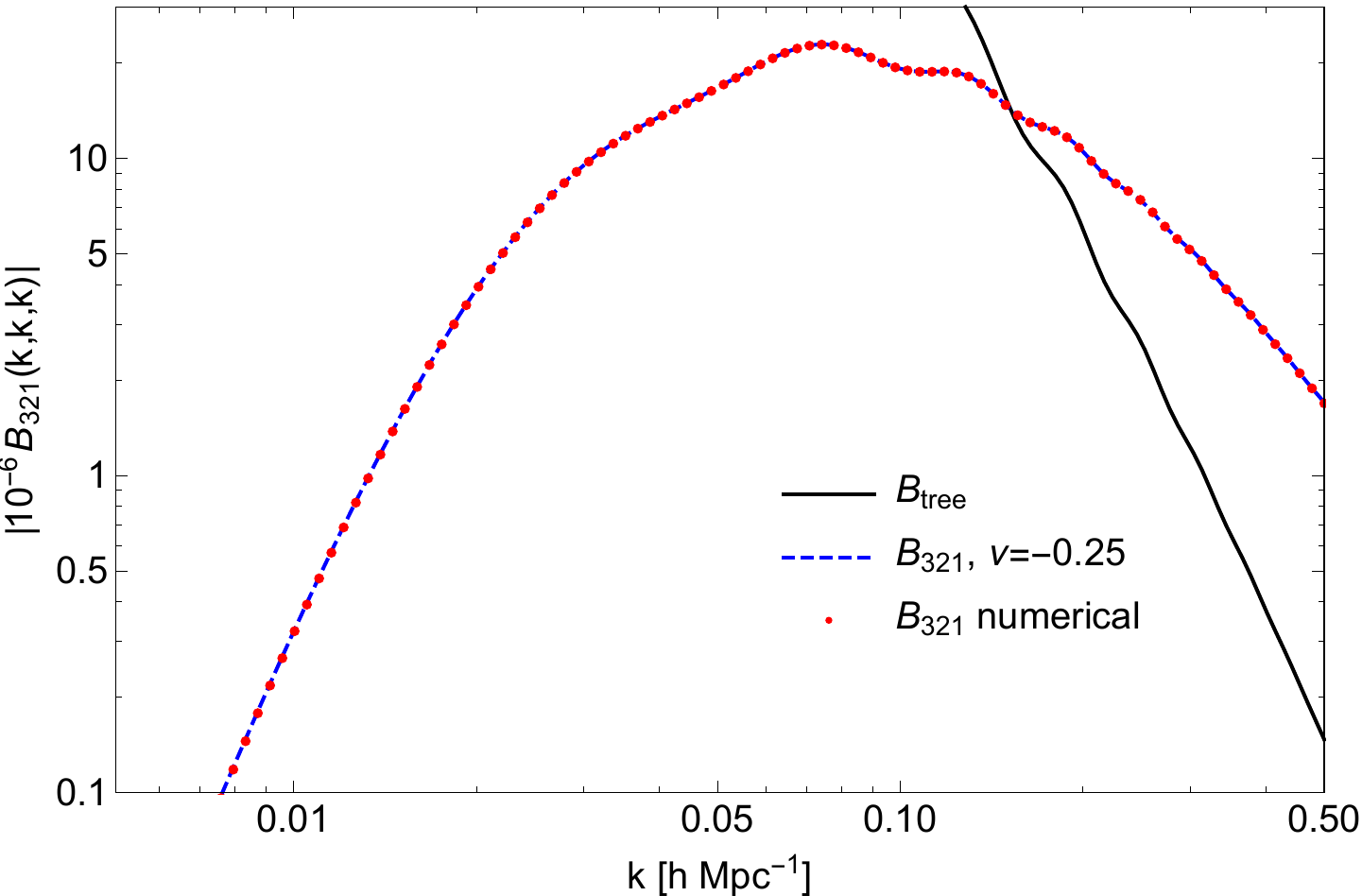}
\end{center}
\caption{Equilateral $B_{321}^I$ diagram as a function of $k$. The calculation is done with bias $\nu=-0.25$ and $N=50$ sampling points for the power spectrum. This is enough to reach the sub-percent precision.}
\label{fig:b321}
\end{figure}
%\vskip 4pt 

\noindent
{\em $B_{321}^I$ term.---}Let us now turn to $B_{321}^I$ diagram. The asymptotic behavior of the $F_3$ kernel in the integrand is the same in the UV and the IR 
\be
F_3(\q,\k_2-\q,-\k_1) \to \frac kq \qquad q\to\infty \quad {\rm and} \quad q\to 0 \;.
\ee
Therefore, the integral is convergent for power laws in the range $-1<\nu<0$. For small negative bias we can use our method without dealing with the possible UV divergences. As in the previous case we can write
\be
\label{eq:B321powers}
\bar B_{321}^I(k_1,k_2,k_3) = k_1^{3} P_{\rm lin}(k_1) \sum_{m_i} M_{321}(\nu_1,\nu_2;x,y) \cdot c_{m_1}k_1^{-2\nu_1}\cdot c_{m_2}k_1^{-2\nu_2} + 5\;{\rm perms}\;,
\ee
where the matrix $M_{321}(\nu_1,\nu_2;x,y)$ is given by
\be
M_{321}(\nu_1,\nu_2;x,y) = 6\sum_{n_i} f_{321}(n_1,n_2,n_3;x,y)\,\J(\cdots;x,y) \;.
\ee
We do not explicitly specify the argument because $\nu_1$ and $\nu_2$ can be at different positions in different terms. The explicit form of the matrix can be found in the {\sf Mathematica} notebook file associated with the preprint of the paper on {\sf arXiv}. 

The same conclusions as in the previous case apply here as well. The matrix $M_{321}$ is cosmology independent and has $N^2$ elements. This makes it numerically much less challenging  than $M_{222}$. In Fig.~\ref{fig:b321} we compare our method for $N=50$ sampling points in the power spectrum with the usual numerical evaluation and find an excellent agreement between the two.
\begin{figure}[h]
\begin{center}
\includegraphics[width=0.7\textwidth]{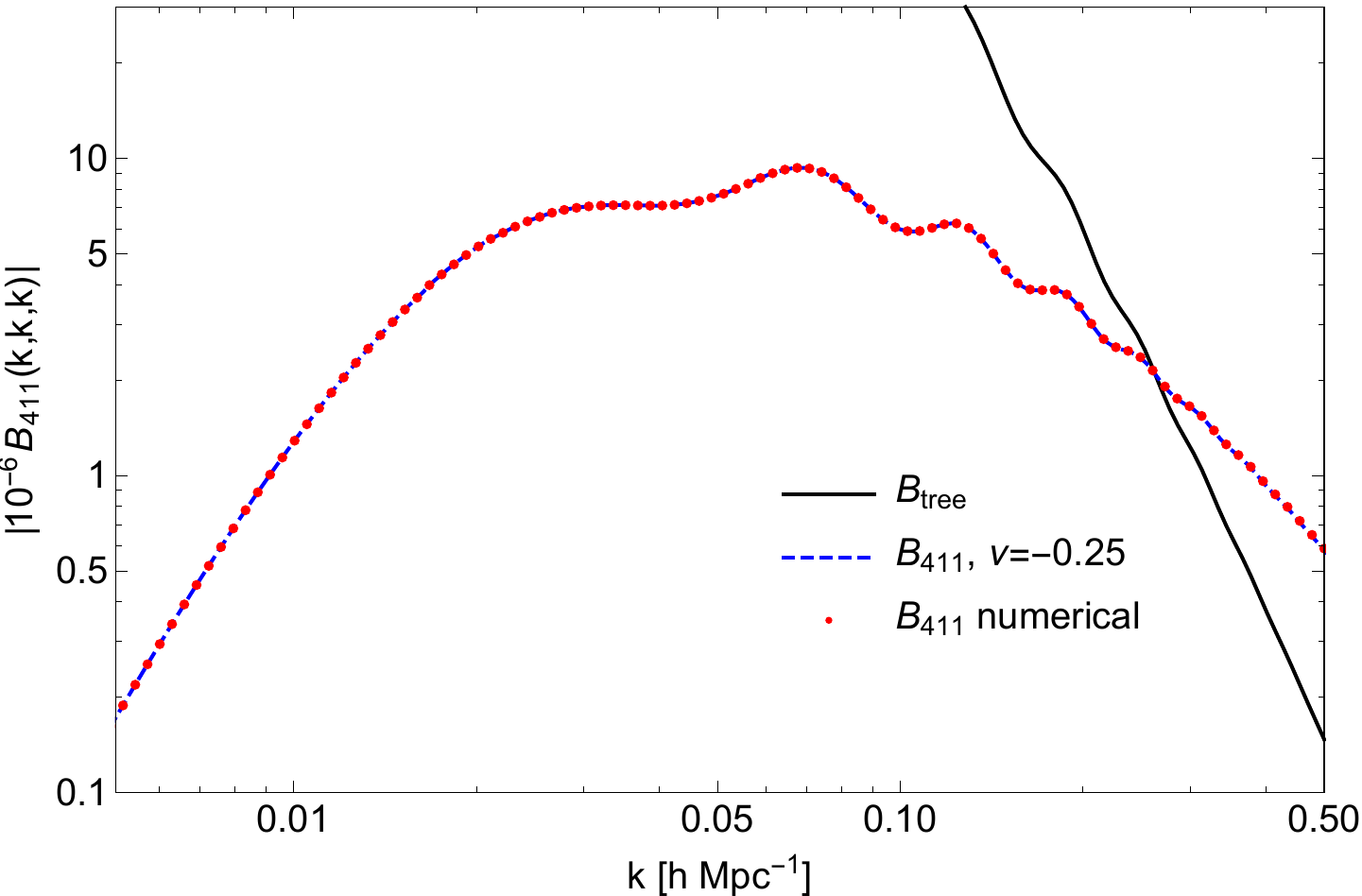}
\end{center}
\caption{Equilateral $B_{411}$ diagram as a function of $k$. The calculation is done with bias $\nu=-0.25$ and $N=50$ sampling points for the power spectrum. This is enough to each the sub-percent precision. }
\label{fig:b411}
\end{figure}
%\vskip 4pt 

\noindent
{\em $B_{411}$ term.---}Unlike for previous diagrams, the $B_{411}$ integral does not converge for any bias. The situation is similar to $P_{13}$ diagram and the problem can be solved in a similar way. For small negative biases the integral is UV divergent. Therefore, to get the usual numerical result, one has to add the UV part of the loop integral which is given by~\cite{Baldauf:2014qfa}
\begin{align}
B_{411}^{\rm UV} =& -\frac{P_{\rm lin} (k_2) P_{\rm lin} (k_3) \sigma_v^2}{226380\,k_2^2k_3^2} \Big( 12409\,k_1^6 + 20085\,k_1^4(k_2^2 + k_3^2) \nonumber \\
&- k_1^2\,(44518\,k_2^4 - 76684\,k_2^2k_3^2 +44518\,k_3^4) + 12024 (k_2^2 - k_3^2)^2 (k_2^2 + k_3^2) \Big)  + 2 \,{\rm perm}. 
\end{align}
As in the case of the power spectrum, dimensional regularization would set these terms to zero. As expected, the structure of the UV part of the $B_{411}$ diagram is such that it can be reabsorbed by the bispectrum counterterms in the EFT approach to LSS~\cite{Baldauf:2014qfa,Angulo:2014tfa}.
%\vskip 4pt 

The regular terms can be organized in a vector $M_{411}$ which has $N$ elements and is cosmology independent. The approximation to the $B_{411}$ diagram can be then written as
\be
\label{eq:B411powers}
\bar B_{411}(k_1,k_2,k_3) = k_1^{3} P_{\rm lin}(k_1) P_{\rm lin}(k_2) \sum_{m} M_{411}(\nu;x,y) \cdot c_{m}k_1^{-2\nu} + 2\;{\rm perms}\;,
\ee
where
\be
M_{411}(\nu;x,y) = 12\sum_{n_i} f_{411}(n_1,n_2,n_3;x,y)\,\J(\cdots;x,y) \;.
\ee
The explicit form of this vector can be found in the {\sf Mathematica} notebook file associated with the preprint of the paper on {\sf arXiv}. In Fig.~\ref{fig:b411} we find an excellent agreement of our method (including $B_{411}^{\rm UV}$ terms) with the usual numerical evaluation.
\begin{figure}[h]
\begin{center}
\includegraphics[width=0.7\textwidth]{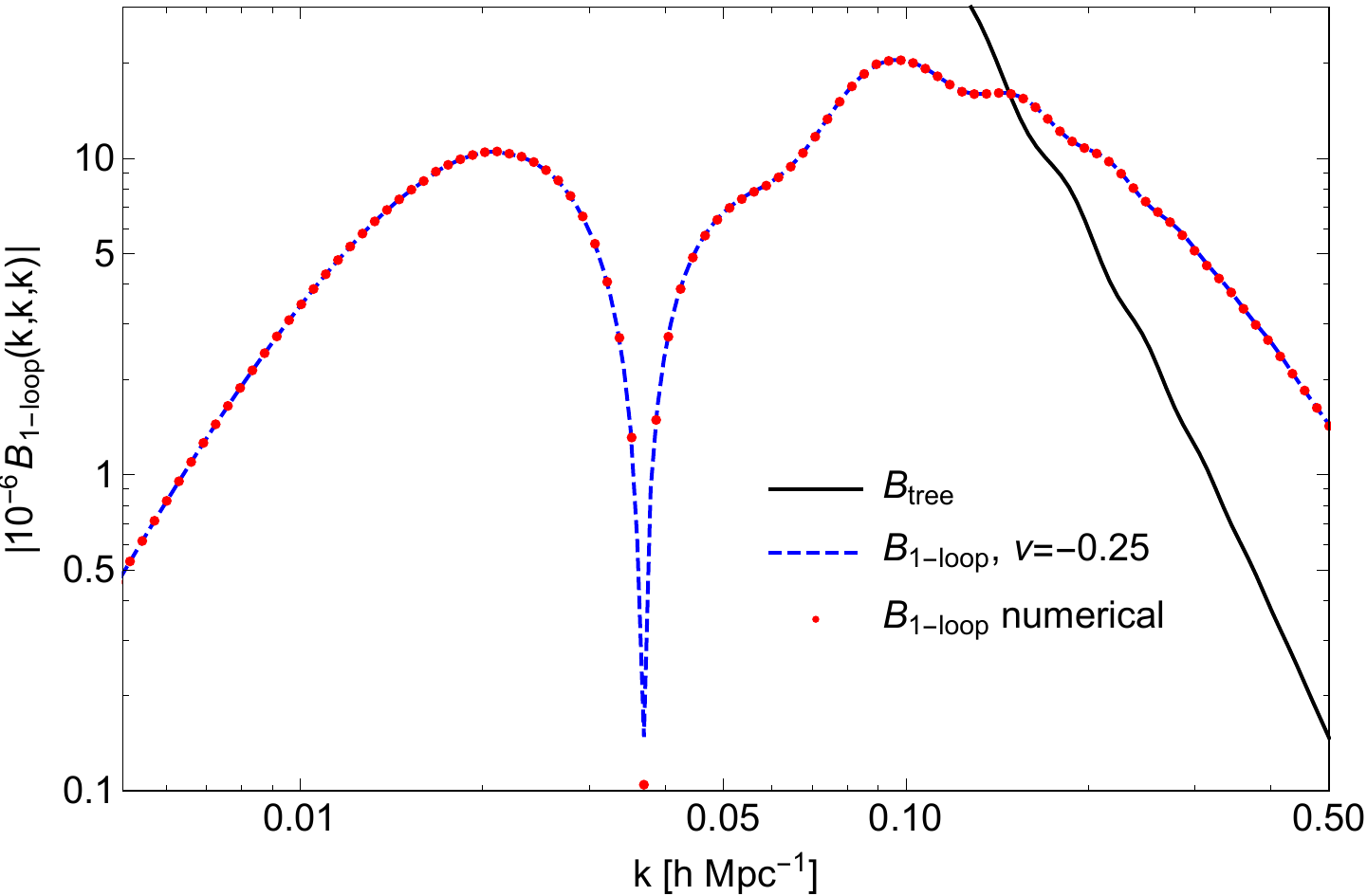}
\end{center}
\caption{Equilateral one-loop bispectrum as a function of $k$. The calculation is done with bias $\nu=-0.25$ and $N=50$ sampling points for the power spectrum. With these parameters it is possible to reach sub-percent precision on all scales. }
\label{fig:b1loop}
\end{figure}
%\vskip 4pt

\noindent
{\em The full one-loop bispectrum.---}In Fig.~\ref{fig:b1loop} we plot the equilateral bispectrum as a function of $k$ and compare our method with the usual numerical result. Given that typically there are no large cancellations between different diagrams, we achieve a similar precision for the full result as for each individual term in the sum. 
%\vskip 4pt 

In summary, for each bispectrum shape given by ratios $x$ and $y$, one has to calculate three matrices $M_{222}$, $M_{321}$ and $M_{411}$. These matrices have $N^3$, $N^2$ and $N$ elements respectively where $N\sim\mathcal O(50)$ is sufficient to achieve sub-percent precision on relevant scales. Computation of these matrices is relatively fast and it can be further optimized using properties of hypergeometric functions. The most practical approach for data analysis is to precompute all matrices and evaluate the one-loop bispectrum with different cosmological parameters as a simple matrix multiplication.

\section{Two-loop Power Spectrum}
\label{sec:2looppower}
\begin{figure}[h]
\begin{center}
\includegraphics[width=1.0\textwidth]{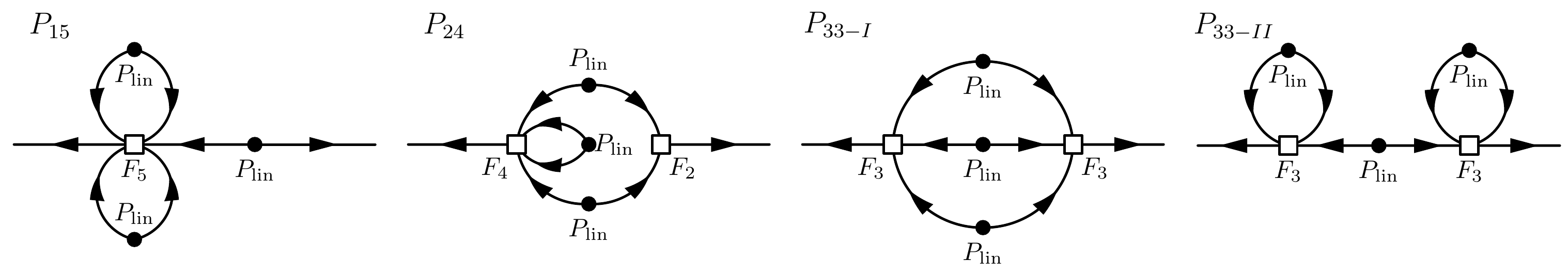}
\end{center}
\caption{Diagrammatic representation of four contributions to the two-loop power spectrum.}
\label{fig:diagrams2LP}
\end{figure}

Now we can turn to the most complicated case of the two-loop power spectrum. There are four different contributions at this order in perturbation theory~\cite{Scoccimarro:1995if, Bernardeau:2001qr} (for the EFTofLSS treatment of the two-loop power spectrum see~\cite{Carrasco:2013sva, Carrasco:2013mua, Baldauf:2015aha, Cataneo:2016suz})
\be
P_{\rm 2-loop}(k,\tau) = D^4(\tau)[P_{33}^{I}(k) + P_{33}^{II}(k) + P_{24}(k) + P_{15}(k) ] \;.
\ee
The explicit form of the four terms in the square brackets is
\be
P_{33}^{I} (k) = 9P_{\rm lin}(k) \int_{\q} F_3(\k,\q,-\q) P_{\rm lin}(q) \int_{\p} F_3(-\k, \p, -\p) P_{\rm lin}(p) \;,
\ee
\be
P_{33}^{II} (k) = 6 \int_{\q} \int_{\p} F_3(\q,\p,\k-\q-\p) F_3(-\q,-\p,\q+\p-\k) P_{\rm lin}(q) P_{\rm lin}(p) P_{\rm lin}(|\k-\q-\p|) \;,
\ee
\be
P_{24} (k) = 24 \int_{\q} \int_{\p} F_2(\q, \k-\q) F_4(\p,-\p,-\q,\q-\k) P_{\rm lin}(q) P_{\rm lin}(p) P_{\rm lin}(|\k-\q|) \;,
\ee
\be
P_{15} (k) = 30 P_{\rm lin}(k) \int_{\q} \int_{\p} F_5(\k,\q,-\q,\p,-\p) P_{\rm lin}(q) P_{\rm lin}(p) \;.
\ee
The corresponding diagrams are shown in Fig.~\ref{fig:diagrams2LP}. In the first contribution, $P_{33}^{I} (k)$, two integrals have the same structure as the $P_{13}(k)$ part of the one-loop calculation. In other cases the integrals are not separable. 

After expanding kernels and linear power spectra in power laws, all terms in the sum can be written in the following form
\begin{align}
\label{eq:Kintdef}
\int_{\q} \frac{1}{q^{2\nu_4} |\k - \q|^{2\nu_5}} \int_{\p} \frac{1}{p^{2\nu_1}  |\k - \p|^{2\nu_2} |\q - \p|^{2\nu_3} } \equiv k^{6-2\nu_{12345}}\, \K(\nu_1,\ldots,\nu_5) \;.
\end{align}
One important point to make is that at most three of five parameters $\nu_1, \ldots,\nu_5$ are generic complex numbers. The reason for this is that there are at most three linear power spectra in the two-loop integrals. The other two parameters must be integers coming from the expansion of kernels. As we are going to see, this simplifies evaluation of some diagrams significantly. 
%\vskip 4pt 

Before we move on, let us notice that the second integral in~\eqref{eq:Kintdef} has identical structure as the one-loop bispectrum. Therefore, choosing the following change of coordinates $x=|\k-\q|^2/k^2$ and $y=q^2/k^2$, the function $\K(\nu_1,\ldots,\nu_5)$ that we are ultimately interested in can be written as follows 
\be
\label{eq:KintrepJ}
\K(\nu_1,\ldots,\nu_5) = \frac{1}{16\pi^2} \int_{x,y} x^{-\nu_5} y^{-\nu_4} \J(\nu_1,\nu_2,\nu_3;x,y) \;.
\ee
We are going to use this equation and the series representation of $\J(\nu_1,\nu_2,\nu_3;x,y)$ to find the explicit formula for $\K(\nu_1,\ldots,\nu_5)$. One difficulty is that the region of integration is rather complicated: $\sqrt{x}+\sqrt{y}\geq 1$ and $|\sqrt{x}-\sqrt{y}|\leq 1$. This reflects the constraints that physical momenta in the two-loop diagram have to satisfy. Before we discuss the explicit expression, we derive some symmetry properties of the function $\K(\nu_1,\ldots,\nu_5)$. These properties will prove useful in practical applications to the full two-loop integral.

\subsection{Symmetries of $\K(\nu_1,\ldots,\nu_5)$ and Recursion Relations}
\label{sec:2looppowersym}
The two-loop diagram of massless theory is known to have a lot of symmetries which translates to many symmetries of the function $\K(\nu_1,\ldots,\nu_5)$. All symmetry transformations were first derived in~\cite{Gorishnii:1984te}. The full symmetry group is $Z_2\times S_6$ and it has $2\times 6! =1440$ elements~\cite{Broadhurst:1986bx,Barfoot:1987kg}. We review here some of the relevant symmetry transformations. To derive them it is enough to use the integral representation
\begin{align}
\label{eq:Kintrep}
\K(\nu_1,\ldots,\nu_5) = \int_{\q} \frac{1}{q^{2\nu_4} |\hat\k - \q|^{2\nu_5}} \int_{\p} \frac{1}{p^{2\nu_1}  |\hat\k - \p|^{2\nu_2} |\q - \p|^{2\nu_3} }  \;,
\end{align}
where $\hat\k$ is the unit vector. Two obvious symmetries are the following. First, exchanging integration variables $\q$ and $\p$ leads to $(\nu_1,\nu_2) \leftrightarrow (\nu_4,\nu_5)$, leaving $\nu_3$ in the same position
\be
\label{eq:Kperm1}
\K(\nu_1, \nu_2, \nu_3, \nu_4, \nu_5) = \K(\nu_4, \nu_5, \nu_3, \nu_1, \nu_2) \;.
\ee
Second, we can simultaneously shift both momenta $\q\to\k-\q$ and $\p\to\k-\p$. Effectively, this produces the simultaneous exchange $(\nu_1 \leftrightarrow \nu_2)$ and $(\nu_4 \leftrightarrow \nu_5)$
\be
\label{eq:Kperm2}
\K(\nu_1, \nu_2, \nu_3, \nu_4, \nu_5) = \K(\nu_2, \nu_1, \nu_3, \nu_5, \nu_4) \;.
\ee
Additional translation formulas can be derived using translation formulas of $\J(\nu_1,\nu_2,\nu_3;x,y)$. Plugging the transformations~\eqref{eq:Jpermutationsym} into the integral representation~\eqref{eq:Kintrep}, it is straightforward to get the following extra identities
\be
\label{eq:Kperm3}
\K(\nu_1, \nu_2, \nu_3, \nu_4, \nu_5) = \K(\nu_3,\nu_1,\nu_2, \nu_5, \nu_6) \;,
\ee
\be
\label{eq:Kperm4}
\K(\nu_1, \nu_2, \nu_3, \nu_4, \nu_5) = \K(\nu_3,\nu_2,\nu_1, \nu_4, \nu_6) \;,
\ee
where $\nu_6\equiv\tfrac 92-\nu_{12345}$. Transformations~\eqref{eq:Kperm1} to~\eqref{eq:Kperm4} form the symmetric group of degree four~$S_4$. This group has $4!=24$ different elements. We recover  these elements by starting from $\K(\nu_1, \nu_2, \nu_3, \nu_4, \nu_5)$ and successively applying identities~\eqref{eq:Kperm1} to~\eqref{eq:Kperm4}. Invoking notation $(1,2,3,4,5)\equiv \K(\nu_1,\ldots,\nu_5)$, then the 24 equivalent functions are
\begin{equation}
\label{eq:KS4group}
\begin{split}
& (1,2,3,4,5) \qquad (2,1,3,5,4) \qquad (4,5,3,1,2) \qquad (5,4,3,2,1) \\
& (3,1,2,5,6) \qquad (1,3,2,6,5) \qquad (5,6,2,3,1) \qquad (6,5,2,1,3) \\
& (3,2,1,4,6) \qquad (2,3,1,6,4) \qquad (4,6,1,3,2) \qquad (6,4,1,2,3) \\
& (3,5,4,1,6) \qquad (5,3,4,6,1) \qquad (1,6,4,3,5) \qquad (6,1,4,5,3) \\
& (3,4,5,2,6) \qquad (4,3,5,6,2) \qquad (2,6,5,3,4) \qquad (6,2,5,4,3) \\
& (5,2,6,4,1) \qquad (2,5,6,1,4) \qquad (4,1,6,5,2) \qquad (1,4,6,2,5) \;.
\end{split}
\end{equation}

Applying inversion (\ref{eq:Jinversionformula}) and star-triangle (\ref{eq:star-triangle}) formulae for $\J(\nu_1,\nu_2,\nu_3;x,y)$ in the integral representation~\eqref{eq:KintrepJ}, provides analogous results for $\K(\nu_1,\ldots,\nu_5)$:
\be
\K(\nu_1,\nu_2,\nu_3,\nu_4,\nu_5) = \K(\nu_2,\nu_1,3-\nu_{123},\nu_{134}- \tfrac32 ,\nu_{235} -\tfrac32 )\;,
\ee
\be
\label{eq:Kstartriangle}
\K(\nu_1,\nu_2,\nu_3,\nu_4,\nu_5) = \tfrac{\Gamma(\tilde\nu_1)}{\Gamma(\nu_1)} \tfrac{\Gamma(\tilde\nu_2)}{\Gamma(\nu_2)} \tfrac{\Gamma(\tilde\nu_3)}{\Gamma(\nu_3)} \tfrac{\Gamma(3-\tilde\nu_{123})}{\Gamma(3-\nu_{123})} \, \K(\tilde\nu_2,\tilde\nu_1,3-\tilde\nu_{123},\nu_4,\nu_5) \;.
\ee
Finally, combining these with translation leads to the following reflection formula
\begin{align}
\label{eq:Kreflection}
\K(\nu_1,&\nu_2,\nu_3,\nu_4,\nu_5) = g\, \K(\tilde\nu_1,\tilde\nu_2,\tilde\nu_3,\tilde\nu_4,\tilde\nu_5) \;,
\end{align}
where the prefactor $g$ is given by
\be
g = \tfrac{\Gamma(\tilde\nu_1)}{\Gamma(\nu_1)} \tfrac{\Gamma(\tilde\nu_2)}{\Gamma(\nu_2)} \tfrac{\Gamma(\tilde\nu_3)}{\Gamma(\nu_3)}
\tfrac{\Gamma(\tilde\nu_4)}{\Gamma(\nu_4)} \tfrac{\Gamma(\tilde\nu_5)}{\Gamma(\nu_5)} \tfrac{\Gamma(\tilde\nu_6)}{\Gamma(\nu_6)} \tfrac{\Gamma(3-\tilde\nu_{123})}{\Gamma(3-\nu_{123})}\tfrac{\Gamma(3-\tilde\nu_{345})}{\Gamma(3-\nu_{345})} \tfrac{\Gamma(\tilde\nu_{235}-3/2)}{\Gamma(\nu_{235}-3/2)} \tfrac{\Gamma(\tilde\nu_{134}-3/2)}{\Gamma(\nu_{134}-3/2)} \;.
\ee 
Applying successively all these transformations one can generate the entire $Z_2\times S_6$ group. However, almost all of these transformations fail to preserve the bias. As  in the case of the one-loop bipsectrum, only  special choices of $\nu$ trivially offer compact relations between functions with different integer parts of the parameters. 
%\vskip 4pt 

Finally, let us write down an example of a recursion relation that $\K(\nu_1,\ldots,\nu_5)$ satisfies. Again, the simplest way to derive these identities is to use the analogous results for $\J(\nu_1,\nu_2,\nu_3;x,y)$. For instance, using recursion relation~\eqref{eq:Jrecursion1} one can immediately write
\begin{align}
\label{eq:Krecursion}
(3-\nu_1-\nu_{123})\, &\K(\nu_1,\nu_2, \nu_3,\nu_4,\nu_5) \nonumber \\
+\, \nu_2\, &\K(\nu_1,\nu_2+1,\nu_3,\nu_4,\nu_5) - \nu_2\, \K(\nu_1-1,\nu_2+1,\nu_3,\nu_4,\nu_5)  \nonumber \\
+ \, \nu_3\, &\K(\nu_1,\nu_2,\nu_3+1,\nu_4-1,\nu_5) - \nu_3\, \K(\nu_1-1,\nu_2,\nu_3+1,\nu_4,\nu_5) = 0  \;.
\end{align}
Other similar expressions can be found using the symmetry transformations discussed above. We will give some more details about the practical application of these equations in the following section. 
\vskip 10 pt

\subsection{Evaluation of $\K(\nu_1,\ldots,\nu_5)$}
\label{sec:2looppowerevaluation}
As we already mentioned, finding an explicit formula for $\K(\nu_1,\ldots,\nu_5)$ is not straightforward. In the integral
\be
\label{eq:Kfullintegral}
\K(\nu_1,\ldots,\nu_5) = \frac{1}{16\pi^2} \int_{x,y} x^{-\nu_5} y^{-\nu_4} \J(\nu_1,\nu_2,\nu_3,x,y) \;,
\ee
the region of integration is complicated and we lack a simple representation of~$\J(\nu_1,\nu_2,\nu_3;x,y)$  convergent in the entire domain. Our strategy to find the full solution is to first concentrate on a part of the region of integration and then use symmetries~\eqref{eq:Jpermutationsym} to find solutions elsewhere. One possibility is to start with the integral
\be
\label{eq:KregD1}
K(\nu_1,\ldots,\nu_5)  \equiv  \int_{\mathcal D_1} x^{-\nu_5} y^{-\nu_4} J(\nu_1,\nu_2,\nu_3,x,y) \;,
\ee
over the following domain $\mathcal D_1 =\{ (x,y)\; |\; \sqrt{x}+\sqrt{y}\geq 1,\; x\leq1,\; y\leq 1 \}$. Given that both $x$ and $y$ are smaller than 1, we can use result~\eqref{eq:Jseries} to evaluate the integral. The details of this derivation are given in Appendix~\ref{app:2-loop}. Here we report only the final formula. We write the result in a form which resembles the series representation of $\J(\nu_1,\nu_2,\nu_3;x,y)$
\begin{align}
\label{eq:Kreg1}
K(\nu_1,&\nu_2,\nu_3,\nu_4,\nu_5)  = \frac{\sec(\pi\nu_{23})}{8\sqrt\pi \Gamma(\nu_1) \Gamma(\nu_2) \Gamma(\nu_3) \Gamma(3-\nu_{123})}  \nonumber \\
& \times \sum_{n=0}^\infty \left[ a_{n}(\nu_1,\nu_2,\nu_3) \; \kappa \left( \tfrac 32-\nu_{235}+n ,-\nu_4, \nu_1+n , \tfrac 32-\nu_2+n, 3-\nu_{23}+2n \right) \right. \nonumber \\
& \qquad \left. -  a_{n}(\tilde\nu_1,\tilde\nu_2,\tilde\nu_3) \;\kappa \left(-\nu_5+n,\tfrac 32-\nu_{134}, \tilde\nu_1+n, \tfrac 32 -\tilde\nu_{2}+n, 3-\tilde\nu_{23}+2n \right) \right] \;.
\end{align} 
The complicated part of the answer is the function $\kappa(\cdots)$ which is given in terms of generalized hypergeometric functions
\begin{align}
\kappa (\alpha,\beta,a,b,c) & = \tfrac{1}{1+\alpha} \left[ \tfrac{1}{1+\beta}\;_3F_2\left(\begin{array}{c}1,a,b; \\2+\beta,c ;\end{array}1\right) \right. - 2\cdot{}_4F_3\left(\begin{array}{c}a,b,1+\beta,\frac{3}{2}+\beta;
\\1-d ,\frac{5}{2}+\gamma,3+\gamma ;\end{array}1\right)
\nonumber \\
& \quad \left. -2\cdot{}_4F_3\left(\begin{array}{c} c-a,c-b,1+\beta+d,\frac32+\beta+d;
\\1+d,d+\gamma+\frac{5}{2},3+\gamma +d;\end{array}1\right) \right] \;,
\end{align}
where $d\equiv c-a-b$ and $\gamma\equiv\alpha+\beta$. However, this is not the end of the story. We have found only one piece of the final answer, which corresponds to the integral over the region $\mathcal D_1$. If one splits the remaining part of domain of integration in~\eqref{eq:Kfullintegral} in the following way (see Fig.~\ref{fig:domain})
\begin{align}
\mathcal D_2 & =\{ (x,y)\; |\; \sqrt{x}-\sqrt{y}\leq 1,\; x\geq1,\; y\leq x \}\;, \\
\mathcal D_3 & =\{ (x,y)\; |\; \sqrt{y}-\sqrt{x}\geq 1,\; y\geq1,\; x\leq y \}\;,
\end{align}
then we find that the integrals over $\mathcal D_2$ and $\mathcal D_3$ can be mapped to an integral of the form~\eqref{eq:KregD1}. The proof is based on symmetries of~$\J(\nu_1,\nu_2,\nu_3;x,y)$ given in eq.~\eqref{eq:Jpermutationsym}. The full solution is then just a sum of three terms given by~\eqref{eq:Kreg1}, with slightly different parameters
\begin{align}
\label{eq:Kthreeparts}
\K(\nu_1,\ldots,\nu_5) = & \frac{1}{16\pi^2} \Big( K(\nu_1,\nu_2,\nu_3,\nu_4,\nu_5) +  K(\nu_1,\nu_3,\nu_2, \nu_6, \nu_5) + K(\nu_2,\nu_3,\nu_1, \nu_6 ,\nu_4) \Big) \;.
\end{align}
Obviously, the final result for $\K(\nu_1,\ldots,\nu_5)$ is very complicated and not very illuminating. Symmetries of the two-loop diagram that we discussed in the previous section are not manifest at all. It is actually quite remarkable that this messy formula satisfies all functional identities of $\K(\nu_1,\ldots,\nu_5)$. This is also an indication that there may exist a much simpler and elegant representation. However, for the time being, it remains elusive. 
%\vskip 4pt 

The difficulties with the result~\eqref{eq:Kreg1} are not only aesthetic but also practical. The biggest problem is that the sum on the r.h.s.~of~\eqref{eq:Kreg1} is not convergent for all values of parameters. Even when it is, sometimes there are big cancellations between different terms in the sum. Summing up many large numbers which eventually leads to a small answer can be numerically quite challenging. On the other hand, there is a large region of parameter space where the sum converges very rapidly. Using many of the symmetries of $\K(\nu_1,\ldots,\nu_5)$ such as~\eqref{eq:KS4group}, it is always possible to evaluate the sum efficiently for any choice of parameters. Understanding the radius of convergence more quantitatively is very important for knowing ahead of time which symmetry transformation to use. Due to complexity of the final answer making some analytic progress is hard and we leave it for the future work.
%\vskip 4pt

Another problem with eq.~\eqref{eq:Kreg1} is the appearance of generalized hypergeometric functions whose argument is equal to 1. At this point the hypergeometric series which is usually used to calculate the function is either divergent or converges very slowly for a generic set of complex parameters. In order to calculate the hypergeometric functions efficiently one can use some of the functional identities such as recursion relations. In all examples when the integral in definition of $\K(\nu_1,\ldots,\nu_5)$ is convergent, the eq.~\eqref{eq:Kthreeparts} agrees with the result of numerical integration. Our power series representation is typically several orders of magnitude faster. 
%\vskip 4pt

So far we considered the general case where the function $\K(\nu_1,\ldots,\nu_5)$ depends on five arbitrary complex numbers. However, as we already pointed out, at least two of $(\nu_1,\ldots,\nu_5)$ are integers which come from the expansion of perturbation theory kernels. In this more specialized case some of the formulas above simplify. There are two different situations that we meet in practice.
%\vskip 4pt

\noindent
{\em One of integer parameters is zero or negative.---}The simplest case is when one of integer parameters is zero. The integral \eqref{eq:Kintdef} becomes a product of two one-loop expressions and the result can be written in terms of gamma functions. For example, let us imagine that $\nu_1=0$. It follows
\begin{align}
\label{eq:Kparameter0}
& \K(0,\nu_2,\ldots,\nu_5) = k^{-6+2\nu_{2345}} \int_{\q} \frac{1}{q^{2\nu_4} |\k - \q|^{2\nu_5} } \int_{\p} \frac{1}{|\k - \p|^{2\nu_2} |\q - \p|^{2\nu_3} } \nonumber \\
& \quad =  k^{-6+2\nu_{2345}} \I(\nu_2,\nu_3) \int_{\q} \frac{ |\k-\q|^{3 - 2 \nu_{23}}}{q^{2\nu_4} |\k - \q|^{2\nu_5} } =  \I(\nu_2,\nu_3)  \I(\nu_4,\nu_{235}-\tfrac32) \;.
\end{align}
The next simplest case is when none of integer parameters is zero, but rather one of them is negative. Let us imagine that $\nu_1=-N$, where $N>0$. In this case the infinite sum in $\J(\nu_1,\nu_2,\nu_3;x,y)$ truncates (see eq.~\eqref{eq:Jtruncates}). The integral \eqref{eq:Kintdef} can again be expressed in terms of gamma functions only. It is straightforward to get
\begin{align}
& \K(-N,\nu_2, \nu_3,\nu_4,\nu_5) = (-1)^{N+1} \frac{\sqrt \pi \sec(\pi\nu_{23})\sec(\pi\nu_3)}{8 \Gamma(\nu_2) \Gamma(\nu_3) \Gamma(3+N-\nu_{23})} \sum_{n=0}^N \sum_{m=0}^{N-n}  \nonumber \\
& \qquad \frac{N!\;(-1)^{m+n} }{(N-m-n)!\,n!\,m!} \frac{\Gamma(\tfrac 32-\nu_2+n+m)}{\Gamma \left(\frac 52-\nu_{23}+ n\right)\Gamma(\nu_{3} - N -\tfrac 12+m)} \, \I(\nu_4-m, \nu_{235}-\tfrac32-n) \;. 
\end{align}
Notice that for $N=0$ this expression reduces to~\eqref{eq:Kparameter0}. This is in agreement with results of~\cite{Schmittfull:2016yqx} (see Appendix F of~\cite{Schmittfull:2016yqx}). In practice, the sums always have at most a few terms. For all diagrams in the two-loop power spectrum $N\leq 5$. The cases in which other parameters are non-positive integers can be easily evaluated using translation formulas~\eqref{eq:KS4group}. 
%\vskip 4pt

Let us point out that in the expansion of the perturbation theory kernels in the $P_{33}$, $P_{24}$ and $P_{15}$ diagrams, most of the terms do have a negative(zero) integer parameter. For example, the expansion of $F_5$ kernel in the $P_{15}$ contribution has several thousand terms. Only ten of them have two positive integer parameters. In other words, the largest part of the two-loop result can be written in terms of gamma functions. Given that in dimensional regularization we do not expect different terms to have very different magnitudes, even neglecting contributions with two positive integer parameters may not affect the result considerably. 
%\vskip 4pt 

\noindent
{\em Both integer parameters are positive.---}Finally, let us discuss the option in which both integer parameters are positive. For the two-loop integral the only possibility is that both of these parameters are equal to one. This comes from the fact that we can have multiple inverse Laplacians in the perturbation theory kernels, but we never have a square of the inverse Laplacian. Under translation formulas~\eqref{eq:KS4group},  all the apparently different cases reduce to the following two cases
\be
\K(1,\nu_2,\nu_3,\nu_4,1) \qquad {\rm and} \qquad \K(1,\nu_2,\nu_3,1,\nu_5) \;.
\ee
Both of these terms can be calculated using eq.~\eqref{eq:Kthreeparts}. It is worth noting that improved numerical stability may be found by considering the related functions provided by the reflection formula~\eqref{eq:Kreflection}
\be
\K(\tfrac12,\tilde\nu_2,\tilde\nu_3,\tilde\nu_4,\tfrac12) \qquad {\rm and} \qquad \K(\tfrac12,\tilde\nu_2,\tilde\nu_3,\tfrac12,\tilde\nu_5) \;.
\ee
%\vskip 4pt 

As in the case of the one-loop power spectrum and the one-loop bispectrum, from the expansion of $F_n$ kernels we get a lot of terms where $\nu_i$ parameters differ just by an integer. Many of those are related by recursion relations. Let us see how the recursion relations look like in the special case when two parameters are equal to one. For example, if $\nu_1=\nu_4=1$, then eq.~\eqref{eq:Krecursion} becomes
\begin{align}
(1-\nu_{23})\, &\K(1,\nu_2, \nu_3,1,\nu_5)
+\, \nu_2\,\K(1,\nu_2+1,\nu_3,1,\nu_5) - \nu_2\,\K(0,\nu_2+1,\nu_3,1,\nu_5) \nonumber \\
+ \, \nu_3\, &\K(1,\nu_2,\nu_3+1,0,\nu_5) - \nu_3\, \K(0,\nu_2,\nu_3+1,1,\nu_5) = 0  \;.
\end{align}
Notice that in three of the five terms one of the arguments is equal to zero. Therefore, they can be written in terms of gamma functions. In this way we get a simple functional identity which relates~$\K(1,\nu_2,\nu_3,1,\nu_5)$ and~$\K(1,\nu_2+1,\nu_3,1,\nu_5)$. When $\nu_1=\nu_5=1$ we get
\begin{align}
(1-\nu_{23})\, &\K(1,\nu_2, \nu_3,\nu_4,1)
+\, \nu_2\, \K(1,\nu_2+1,\nu_3,\nu_4,1) - \nu_2\,\K(0,\nu_2+1,\nu_3,\nu_4,1) \nonumber \\
+ \, \nu_3\, &\K(1,\nu_2,\nu_3+1,\nu_4-1,1) - \nu_3\, \K(0,\nu_2,\nu_3+1,\nu_4,1) = 0  \;.
\end{align}
This equation is slightly more complicated because only two terms have one zero parameter, but it is still very useful. Similar recursion relations can be found exploiting symmetry properties of $\K(\nu_1,\ldots,\nu_5)$. In practice, these relations can reduce the number of terms that one has to evaluate by a factor of a few. 
%\vskip 4pt 

The bottom line is that by using explicit expressions for $\K(\nu_1,\ldots,\nu_5)$, and its symmetry properties, it is possible to calculate all contributions to the two-loop power spectrum. As before, all information can be compressed in three cosmology independent matrices $M_{33}$, $M_{24}$ and $M_{15}$ which correspond to $P^{II}_{33}$, $P_{24}$ and $P_{15}$ diagrams. Evaluation of these matrices is not trivial because of the convergence properties of the series~\eqref{eq:Kreg1}. However, these matrices have to be calculated only once and once they are known, the evaluation of the two-loop power spectrum for any cosmology is just a simple matrix multiplication. These matrices have at most $N^3$ elements where $N\sim\mathcal O(100)$. Therefore, evaluation of the two-loop power spectrum in one $k$ bin is significantly faster than the usual numerical techniques. We leave the implementation and testing of our algorithm for the two-loop power spectrum for future work.

\section{Conclusions}

In this paper we demonstrate the path forward for the efficient computation of higher multiplicity/loop correlation functions in cosmological perturbation theory. Our starting point is similar to recent proposals for fast evaluation of the one-loop power spectrum \cite{McEwen:2016fjn,Schmittfull:2016jsw} and it is based on representing the linear power spectrum as a sum of complex power laws. However, our implementation and generalization to higher order correlators is different. When comparison is possible, all methods agree. 
%\vskip 4pt

We mainly focus on deriving relevant analytic expressions for the one-loop and the two-loop power spectrum and the one-loop bispectrum. All one-loop diagrams evaluated using our method are in excellent agreement with the usual numerical results. We leave writing a dedicated code for the two-loop power spectrum (and possibly higher order correlation functions) for future work. 
%\vskip 4pt

Our method splits the computation of loop diagrams in two parts. The first, more ``difficult'' part is related to solving momentum integrals for power-law power spectra and it is cosmology independent. The second part is a simple matrix multiplication which evaluates the loops for a $\Lambda$CDM-like cosmology. The matrices can be precomputed, they are cosmology independent and they are relatively small. For example, for the two-loop power spectrum, the largest matrix has $N^3$ elements, where $N\sim\mathcal O(100)$. The number of operations needed for evaluation of the power spectrum or the bispectrum is significantly smaller than using direct numerical integration. Furthermore, the same building blocks used to calculate dark matter correlation functions can be also used for correlators of biased tracers. There are no fundamental obstacles in applying our method in redshift space as well. 
%\vskip 4pt

One interesting aspect of the method described in this paper is that it relies on evaluation of loop integrals that are formally identical to those of a massless QFT. This is a new bridge between cosmology and particle physics and the full potential of this connection is still to be explored. This remains the major direction for future work. One hope is that many developments in the theory of scattering amplitudes will prove useful for going beyond the lowest order statistics discussed in this paper. The first step in this direction is a practical analytic formula for the one-loop trispectrum. In principle, that would allow the calculation of the three-loop power spectrum or the two-loop bispectrum. In practice, following procedure described in this paper may turn out to be too difficult or impractical. After all, the trispectrum is a function of six variables, which makes it much more complicated than examples we considered so far. 
%\vskip 4pt 

However, there are many alternative representations of loop integrals that may be more useful for higher loop diagrams. In this paper we have insisted on finding well-behaved power series representations for functions such as $\I(\nu_1,\nu_2)$ or $\J(\nu_1,\nu_2,\nu_3;x,y)$. Alternative ways to evaluate these integrals include numerical integration using Mellin-Barnes representation of loop integrals (see for instance~\cite{Czakon:2005rk}), projecting onto a basis of other known higher-loop integrals as in~\cite{Johansson:2012sf}, or solving numerically partial differential equations that the loop integrals satisfy~\cite{Henn:2014qga}.
%\vskip 4pt 

In some important situations things simplify. One such example is the one-loop covariance of the power spectrum. Given that there are only two independent vectors $\k_1$ and $\k_2$, this special case of the one-loop four-point function depends only on three variables (before integrating over the angle between $\k_1$ and $\k_2$). Furthermore, a set of diagrams in the one-loop covariance matrix which give the largest contribution to the final answer (see~\cite{Mohammed:2016sre,Barreira:2017kxd}) have the same structure as the one-loop bispectrum. These diagrams can be easily calculated using our function $\J(\nu_1,\nu_2,\nu_3;x,y)$. We leave application of our method to the covariance matrix and more generally one-loop four point function for future work.
%\vskip 4pt 

At the end, let us stress that the idea of representing the $\Lambda$CDM-like cosmology as a set of power-law universes can be also very useful outside the context of PT. One can benefit form this decomposition whenever some numerically heavy integral has a simple solution for a power-law universe. One example of this kind is projection of the power spectrum or the bispectrum on the sky, which is difficult due to many integrals over highly oscillatory spherical Bessel functions. It was shown in~\cite{Assassi:2017lea} that decomposition~\eqref{eq:fftlog} can be used to find the solutions of these integrals very accurately and efficiently. It would be interesting to think of other similar applications in the future. 
\vskip 10 pt

\section*{Acknowledgments}
We would like to thank Nima Arkani-Hamed, Valentin Assassi, Diego Blas, Jonathan Blazek, Paolo Creminelli, Guido D'Amico, Chris Hirata, Lam Hui, Mikhail Ivanov, David Kosower, Marcel Schmittfull, Roman Scoccimarro, Leonardo Senatore, Sergey Sibiryakov, Kris Sigurdson, Zachary Slepian, Gabriele Trevisan and Zvonimir Vlah for many useful discussions.  J.~J.~M.~C. is supported by the
European Research Council under ERC-STG-639729, {\it preQFT: Strategic
  Predictions for Quantum Field Theories}. M.S.~gratefully acknowledges support from the Institute for Advanced Study and the Raymond and Beverly Sackler Foundation. M.Z.~is supported by NSF grants AST-1409709 and PHY-1521097 and by the Canadian Institute for Advanced Research (CIFAR) program on Gravity and the Extreme Universe.

\appendix

\section{\label{app:Hyper}Hypergeometric Functions}
The hypergeometric function $\,_2F_1(a,b,c,z)$ is usually defined as a solution of Euler's hypergeometric equation:
\beq
z(1-z)\,f''(z)+\big(c-(a+b+1)z\big)\,f'(z) -ab\,f(z) = 0\;,
\eeq
where $a$, $b$ and $c$ are arbitrary complex numbers. The hypergeometric function has the power series representation:
\beq
\,_2F_1(a,b,c,z) \ = \ \frac{\Gamma(c)}{\Gamma(a)\Gamma(b)}\sum_{n=0}^\infty \frac{\Gamma(a+n)\Gamma(b+n)}{\Gamma(c+n)n!} z^n \;,
\label{eq:2F1_power}
\eeq
which is convergent inside the unit circle in the complex plane $|z|<1$. This power series can be used for numerical evaluation. The series is convergent at the point $z=1$ only when the parameters satisfy~${\rm Re}(c-a-b)>0$. It should be stressed that the convergence sometimes may be slow or the series has large cancellations, particularly for parameters with large imaginary parts. In order to avoid such issues or evaluate the hypergeometric function outside the unit disc, one can use many functional identities. For example, one such identity is 
\begin{align}
\,_2F_1(a,b,c,1-z) &\ =  \ \frac{\Gamma(c) \Gamma(c - a - b)}{\Gamma(c - a) \Gamma(c - b)} \,_2F_1(a, b, a + b - c + 1, z) \nonumber \\
& \qquad +\ \frac{\Gamma(c) \Gamma(a + b - c)}{\Gamma(a) \Gamma(b)} z^{c - a - b} \,_2F_1(c - a, c - b, 1 - a - b + c, z) \;,
\label{eq:zto(1-z)}
\end{align}
which maps points close to $|z|=1$ to a region around $z=0$ where the series converges rapidly. Outside the unit disc the hypergeometric function can be calculated using 
\beq
{}_2F_1(a,b,c,1/z)\ =\  \frac{\Gamma(b - a) \Gamma(c)}{\Gamma(b) \Gamma(c - a)} (-z)^a\,_2F_1(a, a - c + 1, a - b + 1, z) \; + \; (a\leftrightarrow b)\;.
\label{eq:zto1/z}
\eeq
These two identities are sufficient to evaluate the hypergeometric function in the entire complex plane.
%\vskip 4pt 

It is possible to generalize the basic hypergeometric series~(\ref{eq:2F1_power}) and use it to define generalized hypergeometric functions
\beq
{}_pF_q\left(\begin{array}{c}
a_1\,,\,a_2\,,\ldots,\,a_p\\
b_1\,,\,b_2\,,\ldots,b_q
\end{array};\, z\, 
\right) \ \equiv \ \frac{\Gamma(b_1)\cdots\Gamma(b_q)}{\Gamma(a_1)\cdots \Gamma(a_p)}\sum_{n=0}^\infty \frac{\Gamma(a_1+n)\cdots \Gamma(a_p+n)}{\Gamma(b_1+n)\cdots \Gamma(b_q+n)} \frac{z^n}{n!} \;,
\label{eq:pFq_power}
\eeq
where $p$ and $q$ are positive integers. In this paper we use two generalized hypergeometric functions ${}_4F_3$ and ${}_3F_2$. In these cases when $p=q+1$ the generalized hypergeometric series~(\ref{eq:pFq_power}) converges for $|z|<1$. A the point is $z=1$ which is of special interest in evaluation of the two-loop power spectrum the series converges only when~${\rm Re}(b_1+\cdots+b_q-a_1-\cdots a_{q+1})>0$. This condition is not always satisfied in practice. One simple way out is to use recursion relations which increase the real part of one of $b_i$ until the series becomes convergent.
%\vskip 4pt  

\section{\label{app:Tables}Explicit form of the $M_{222}$ matrix}
In this appendix we give the explicit form of the $M_{222}$ matrix. The starting point are three $F_2$ kernels in the $B_{222}$ diagram
\be
8 F_2 (\q,\k_1-\q) F_2(\k_1-\q,\k_2+\q) F_2(\k_2+\q,-\q) \;.
\ee
Expanding this expression in powers of $q^2$, $|\k_1-\q|^2$ and $|\k_2+\q|^2$ for a single set of parameters $(\nu_1,\nu_2,\nu_3)$ and factoring out $k_1$ dependence we get a sum which can be rearranged in the following way
\begin{align}
&M_{222} = \tfrac{25x^2}{1372}\J_{-2,2,2} -\tfrac{15 (y+1) x^2}{1372}\J_{-1,2,2} -\tfrac{\left(10 y^2-9 y+10\right)x^2}{1372} \J_{0,2,2} +\tfrac{3x^2y (y+1)}{686} \J_{1,2,2} +\tfrac{x^2y^2}{343} \J_{2,2,2} +\tfrac{75x}{2744}\J_{-2,1,2} \nonumber \\
& +\tfrac{75x}{2744}\J_{-2,2,1}-\tfrac{5 x(20 x+9 y+9)}{2744} \J_{-1,1,2} -\tfrac{5x(20x+9y+9)}{2744}\J_{-1,2,1} +\tfrac{3x\left(-10 y^2+9 y+10 x (2 y-1)-10\right)}{2744}\J_{0,1,2} \nonumber \\
&  -\tfrac{3x\left(10 y^2-9 y+10 x (y-2)+10\right)}{2744}\J_{0,2,1} +\tfrac{x y (20 y x+9 x+9 y+9)}{1372}\J_{1,1,2} +\tfrac{x(9 y (y+1)+x (9 y+20))}{1372} \J_{1,2,1} + \tfrac{3x(x+1) y^2}{686} \J_{2,1,2} \nonumber \\
& +\tfrac{3xy(x+y)}{686}\J_{2,2,1} -\tfrac{125}{2744}\J_{-2,0,2} + \tfrac{125}{1372}\J_{-2,1,1} - \tfrac{125}{2744}\J_{-2,2,0} + \tfrac{125}{1372} \J_{-1,-1,2} - \tfrac{125}{1372}\J_{-1,0,1} -\tfrac{75 (2 x-y-1)}{2744}\J_{-1,0,2} \nonumber \\
& -\tfrac{125}{1372}\J_{-1,1,0} -\tfrac{75 (2 x+y+1)}{1372}\J_{-1,1,1} + \tfrac{125}{1372}\J_{-1,2,-1} - \tfrac{75 (2 x-y-1)}{2744}\J_{-1,2,0} - \tfrac{125}{2744}\J_{0,-2,2} - \tfrac{125}{1372}\J_{0,-1,1} \nonumber \\
& + \tfrac{75(x-2y+1)}{2744}\J_{0,-1,2} + \tfrac{375}{1372}\J_{0,-1,2} + \tfrac{75(5x+5y-4)}{2744}\J_{0,0,1} + \tfrac{5 \left(10 x^2+9 (2 y-1) x+10 y^2-9 y+10\right)}{2744}\J_{0,0,2} - \tfrac{125}{1372}\J_{0,1,-1} \nonumber \\
& + \tfrac{75 (5 x-4 y+5)}{2744}\J_{0,1,0} + \tfrac{5 \left(40 x^2+9 (y+1) x-20 y^2+18y-20\right)}{2744}\J_{0,1,1} -\tfrac{125}{2744}\J_{0,2,-2} + \tfrac{75 (x+y-2)}{2744}\J_{0,2,-1} \nonumber \\
& + \tfrac{5\left(10 x^2-9 (y-2) x+10 y^2-9 y+10\right)}{2744}\J_{0,2,0} + \tfrac{125}{1372}\J_{1,-2,1} + \tfrac{75 y}{2744}\J_{1,-2,2} - \tfrac{125}{1372}\J_{1,-1,0} - \tfrac{75 (x+2 y+1)}{1372}\J_{1,-1,1} \nonumber \\
& - \tfrac{5y(9x+20y+9)}{2744}\J_{1,-1,2} -\tfrac{125}{1372}\J_{1,0,-1} - \tfrac{75(4 x-5 (y+1))}{2744}\J_{1,0,0} - \tfrac{5 \left(20x^2-9 (y+2) x-40 y^2-9 y+20\right)}{2744}\J_{1,0,1} \nonumber \\
& + \tfrac{3 y \left(-10 x^2+(20 y+9) x-10 (y+1)\right)}{2744}\J_{1,0,2} + \tfrac{125}{1372}\J_{1,1,-2} - \tfrac{75(x+y+2)}{1372} \J_{1,1,-1} - \tfrac{5\left(20 x^2-9 (2
   y+1) x+20 y^2-9 y-40\right)}{2744}\J_{1,1,0} \nonumber \\
& + \tfrac{3\left(10 (y+1) x^2+\left(10 y^2+9 y+10\right)x+10y (y+1)\right)}{1372}\J_{1,1,1} + \tfrac{75}{2744}\J_{1,2,-2} - \tfrac{5 (9 x+9 y+20)}{2744}\J_{1,2,-1} \nonumber \\
& - \tfrac{3\left(10 x^2-(9 y+20)x+10y(y+1)\right)}{2744}\J_{1,2,0} - \tfrac{125}{2744}\J_{2,-2,0} + \tfrac{75y}{2744}\J_{2,-2,1} + \tfrac{25 y^2}{1372}\J_{2,-2,2} + \tfrac{125}{1372}\J_{2,-1,-1}  \nonumber \\
& + \tfrac{75 (x-2 y+1)}{2744}\J_{2,-1,0} - \tfrac{5 y (9 x+20 y+9)}{2744}\J_{2,-1,1} - \tfrac{15(x+1)y^2}{1372}\J_{2,-1,2} - \tfrac{125}{2744}\J_{2,0,-2} + \tfrac{75 (x+y-2)}{2744}\J_{2,0,-1}  \nonumber \\
& + \tfrac{5 \left(10 x^2-9 (y+1) x+2 \left(5 y^2+9y +5\right)\right)}{2744}\J_{2,0,0} - \tfrac{3 y \left(10 x^2+(10 y-9) x-20 y+10\right)}{2744}\J_{2,0,1} -\tfrac{\left(10 x^2-9 x+10\right) y^2}{1372}\J_{2,0,2} \nonumber \\
& + \tfrac{75}{2744}\J_{2,1,-2} - \tfrac{5(9x+9y+20)}{2744}\J_{2,1,-1} - \tfrac{3\left(10 x^2+(10-9 y) x+10 (y-2) y\right)}{2744}\J_{2,1,0} + \tfrac{y\left(9 x^2+9 (y+1) x+20 y\right)}{1372}\J_{2,1,1} \nonumber \\
& + \tfrac{25}{1372}\J_{2,2,-2} - \tfrac{15 (x+y)}{1372}\J_{2,2,-1} - \tfrac{\left(10 x^2-9 y x+10 y^2\right)}{1372}\J_{2,2,0}\;.
\end{align}
We are using shorten notation in which $\J_{n_1,n_2,n_3}\equiv\J(\nu_1+n_1,\nu_2+n_2,\nu_3+n_3;x,y)$. Coefficients $f_{222}$ can be easily read off form this expression. Notice that there are 72 terms in the sum, but not all of them are independent. Using recursion relations~\eqref{eq:Jrecursion} one can further reduce this expression to a sum of 38 different $\J_{n_1,n_2,n_3}$ functions. We do not write this sum explicitly because the coefficients multiplying $\J_{n_1,n_2,n_3}$ functions become too cumbersome. Nevertheless, these new coefficients are still only rational functions that depend on $x$, $y$, $\nu_1$, $\nu_2$ and $\nu_3$ and application of recursion relations effectively reduces the cost of evaluating the $B_{222}$ diagram by roughly a factor of 2. 

\section{\label{app:J}Derivation of $\J(\nu_1,\nu_2, \nu_3 ;x,y)$} 
Let us begin with the usual Feynman parametrization
\begin{align}
&\int_{\q} \frac{1}{q^{2\nu_1}|\k_1-\q|^{2\nu_2} |\k_2+\q|^{2\nu_3} } =\frac{\Gamma(\nu_{123})}{\Gamma(\nu_1) \Gamma(\nu_2) \Gamma(\nu_3)} \nonumber \\
& \quad \times \int_0^1 du_1 \int_0^1 du_2 \int_0^1 du_3 \int_{\q} \; \frac{u_1^{\nu_1-1}u_2^{\nu_2-1}u_3^{\nu_3-1}\delta^\text{(D)}(1-u_1-u_2-u_3)}{\left(u_1q^2 +u_2|\k_1-\q|^2 + u_3 |\k_2+\q|^2 \right)^{\nu_{123}}} \nonumber \\
& = \frac{\Gamma(\nu_{123})}{\Gamma(\nu_1) \Gamma(\nu_2) \Gamma(\nu_3)}  \int_0^1 du_1 \int_0^{1-u_1} du_2 \int_{\q} \;\frac{u_1^{\nu_1-1}u_2^{\nu_2-1}(1-u_1-u_2)^{\nu_3-1}}{\left(u_1q^2 +u_2|\k_1-\q|^2 + (1-u_1-u_2) |\k_2+\q|^2 \right)^{\nu_{123}}} \;.
\end{align}
Next, we do the following change of variables: $u_1=uv$ and $u_2=(1-u)v$. This transforms $(1-u_1-u_2)$ into $v$ and now both integrals in $u$ and $v$ have the same boundaries $[0,1]$
\begin{align}
& \int_{\q} \frac{1}{q^{2\nu_1}|\k_1-\q|^{2\nu_2} |\k_2+\q|^{2\nu_3} } = \frac{\Gamma(\nu_{123})}{\Gamma(\nu_1)\Gamma(\nu_2)\Gamma(\nu_3)} \nonumber \\
& \qquad\qquad \times \int_{0}^1 du \int_0^1 dv \int_{\q} \;\frac{u^{\nu_1-1}(1-u)^{\nu_2-1}v^{\nu_{12}-1}(1-v)^{\nu_3-1}}{\left( uvq^2 +(1-u)v|\k_1-\q|^2 +(1-v)|\k_2+\q|^2 \right)^{\nu_{123}}} \;.
\end{align}
At this point the momentum integral can be done easily. In the denominator we first complete the square
\begin{align}
& uvq^2 +(1-u)v|\k_1-\q|^2 +(1-v)|\k_2+\q|^2 \nonumber \\
& \quad = \left( \q - (1-u) v\k_1 + (1-v)\k_2 \right)^2 + v \left( u v (1-u) k_1^2 + u(1-v)k_2^2 + (1-u) (1-v) k_3^2  \right) \;,
\end{align}
and use the following identity to do the integral in $\q$
\be
\int_{\q} \frac{1}{(q^2 +m^2)^{\nu_{123}}} = \frac1{8\pi^{3/2}} \frac{\Gamma\left(\nu_{123}-\frac32 \right)}{\Gamma(\nu_{123})} \frac1{(m^2)^{\nu_{123}-3/2}} \;.
\ee
The expression for one-loop bispectrum simplifies and we are left with two integrals in $u$ and $v$
\begin{align}
& \int_{\q} \frac{1}{q^{2\nu_1}|\k_1-\q|^{2\nu_2} |\k_2+\q|^{2\nu_3} } = \frac{k_1^{3-2\nu_{123}}}{8\pi^{3/2}} \frac{\Gamma\left( \nu_{123}-\frac32 \right)}{\Gamma(\nu_1)\Gamma(\nu_2)\Gamma(\nu_3)}  \nonumber \\
& \qquad\qquad \times \int_{0}^1 du \int_0^1 dv \frac{u^{\nu_1-1}(1-u)^{\nu_2-1}v^{1/2-\nu_3}(1-v)^{\nu_3-1}}{\left( u v (1-u) + u(1-v) y + (1-u) (1-v) x \right)^{\nu_{123}-3/2}} \;,
\end{align}
from which we can read off $\J(\nu_1,\nu_2, \nu_3 ;x,y)$
\begin{align}
\J(\nu_1,\nu_2, \nu_3 ;x,y) & = \frac{1}{8\pi^{3/2}} \frac{\Gamma\left(\nu_{123}-\frac32 \right)}{\Gamma(\nu_1)\Gamma(\nu_2)\Gamma(\nu_3)} \nonumber \\
& \quad \times \int_{0}^1 du \int_0^1 dv \frac{u^{\nu_1-1}(1-u)^{\nu_2-1}v^{1/2-\nu_3}(1-v)^{\nu_3-1}}{\left( u v (1-u) + u(1-v) y + (1-u) (1-v) x \right)^{\nu_{123}-3/2}} \;.
\end{align}
Notice that the denominator is linear in $v$ and that the integral in $v$ is nothing but the hypergeometric function
\begin{align}
&\J(\nu_1,\nu_2,\nu_3 ;x,y) = \frac{\Gamma\left(\frac32-\nu_3\right) \Gamma \left( \nu_{123}-\frac32 \right)} {4\pi^2\Gamma(\nu_1) \Gamma(\nu_2)} \int_0^1 du\;u^{\nu_1-1}(1-u)^{\nu_2-1} \nonumber \\
&\quad \times (x(1-u)+y u)^{3/2-\nu_{123}}\;_2F_1\left(\frac 32 - \nu_3, \nu_{123}-\frac 32, \frac 32, 1- \frac{u(1-u)}{x(1-u)+y u}\right) \;.
\end{align}
At this point it is useful to transform this expression using~\eqref{eq:zto(1-z)}
\begin{align}
&\J(\nu_1,\nu_2,\nu_3;x,y) = \frac{\Gamma\left(\frac32-\nu_3\right) \Gamma \left( \nu_{123}-\frac32 \right)} {4\pi^2\Gamma(\nu_1) \Gamma(\nu_2)}  \int_0^1 du\; u^{\nu_1-1}(1-u)^{\nu_2-1}(x(1-u)+y u)^{3/2-\nu_{123}} \nonumber \\
& \quad \left[ \frac{\Gamma\left( \frac32 \right) \Gamma\left(\nu_{12}-\frac32 \right)}{\Gamma\left( \frac32-\nu_3 \right) \Gamma\left( \nu_{123}-\frac 32 \right)} \frac{u^{3/2-\nu_{12}}(1-u)^{3/2-\nu_{12}}}{(x(1-u)+y u)^{3/2-\nu_{12}}} \;_2F_1\left(\nu_3, 3-\nu_{123}, \frac 52-\nu_{12}, \frac{u(1-u)}{x(1-u)+y u}\right) \right. \nonumber \\
& \quad \left. + \frac{\Gamma\left( \frac32 \right) \Gamma\left( \frac32- \nu_{12} \right)}{\Gamma(\nu_3)\Gamma(3-\nu_{123})} \;_2F_1\left(\frac 32 - \nu_3, \nu_{123}-\frac 32, \nu_{12}- \frac 12, \frac{u(1-u)}{x(1-u)+y u}\right)\right] \;.
\end{align}
The reason is that $0\leq u(1-u)/(x(1-u)+y u)\leq1$ for any $x$ and $y$, and one can use the power series representation of hypergeometric functions in order to solve the integral in $u$.
Notice that this power series keeps the integral as simple as  possible, because only powers or $u$, $(1-u)$ and $(x(1-u)+y u)$ appear in the expression. Simplifying the gamma functions we get
\begin{align}
\J(\nu_1,\nu_2,& \nu_3;x,y) = \frac{\sec(\pi\nu_{12})} {8\sqrt\pi \Gamma(\nu_1) \Gamma(\nu_2) \Gamma(\nu_3)\Gamma(3-\nu_{123}) }  \nonumber \\
& \quad \left[ \sum_{n=0}^\infty \frac{\Gamma(\nu_3+n) \Gamma(3-\nu_{123}+n)}{\Gamma\left(\frac 52-\nu_{12}+n \right)n!} \int_0^1 du\; \frac{u^{1/2-\nu_{2}+n}(1-u)^{1/2-\nu_{1}+n}}{(x(1-u)+y u)^{\nu_{3}+n}} \right. \nonumber \\
& \quad \left. - \sum_{n=0}^\infty  \frac{\Gamma\left(\frac 32 - \nu_3+n \right) \Gamma\left( \nu_{123}-\frac 32+n\right)}{\Gamma\left( \nu_{12}- \frac 12+n \right)n!} \int_0^1 du\; \frac{u^{\nu_1-1+n}(1-u)^{\nu_2-1+n}}{(x(1-u)+y u)^{\nu_{123}-3/2+n}}   \right] \;.
\end{align}
The integration in $u$ leads to another hypergeometric function. The result can be written in the following way
\begin{align}
\J(\nu_1, & \nu_2,\nu_3;x,y) = \frac{\sec(\pi\nu_{12})} {8\sqrt\pi\Gamma(\nu_1) \Gamma(\nu_2) \Gamma(\nu_3)\Gamma(3-\nu_{123}) }  \nonumber \\
& \; \left[ \sum_{n=0}^\infty b_n(\nu_1,\nu_2,\nu_3)\cdot x^{-\nu_3-n} \;_2F_1\left(\frac 32 - \nu_2+n, \nu_{3}+n, 3-\nu_{12}+2n, 1-\frac yx \right) \right. \nonumber \\
& \quad \left. - \sum_{n=0}^\infty b_n(\tilde\nu_1,\tilde\nu_2,\tilde\nu_3)\cdot  x^{-\nu_{2}-n} y^{3/2-\nu_{13}} \;_2F_1\left(\frac32-\nu_3+n, \nu_2+n, \nu_{12}+2n, 1-\frac yx \right)  \right] \;,
\end{align}
where the coefficients $b_n$ are given by
\be
b_n(\nu_1,\nu_2,\nu_3) = \frac{\Gamma(\nu_3+n) \Gamma(3-\nu_{123}+n)}{\Gamma\left(\frac 52-\nu_{12}+n \right)n!} \frac{\Gamma\left(\frac 32-\nu_{1}+n \right) \Gamma\left(\frac 32-\nu_{2}+n \right)}{\Gamma\left(3-\nu_{12}+2n \right)} \;.
\ee
One last step is to use the identity 
\be
\J\left(\nu_1,\nu_2,\nu_3;x,y\right)=x^{3/2-\nu_{123}} \J\left(\nu_3,\nu_2,\nu_1;\frac{1}{x}, \frac{y}{x}\right)\;,
\ee
in order to bring the result to its final form
\begin{align}
\J(\nu_1,\nu_2, & \nu_3;x,y) = \frac{\sec(\pi\nu_{23})}{8\sqrt \pi \Gamma(\nu_1) \Gamma(\nu_2) \Gamma(\nu_3) \Gamma(3-\nu_{123})}\nonumber \\
&\sum_{n=0}^\infty  \left[ x^{3/2-\nu_{23}} \cdot b_{n}(\nu_3,\nu_2,\nu_1) \; x^{n}\;_2F_1\left(\nu_1+n, \frac 32-\nu_2+n, 3-\nu_{23}+2n, 1-y \right) \right. \nonumber \\
& \quad \left. -y^{3/2-\nu_{13}} \cdot b_{n}(\tilde\nu_3,\tilde\nu_2,\tilde\nu_1) \; x^{n}\;_2F_1\left(  \nu_2+n, \frac 32 -\nu_{1}+n, \nu_{23}+2n, 1-y  \right) \right] \;.
\end{align}
This precisely matches eq.~\eqref{eq:Jseries} where $a_n(\nu_1,\nu_2,\nu_3) = b_n(\nu_3,\nu_2,\nu_1)$.

\section{\label{app:2-loop}Derivation of $\K(\nu_1,\ldots,\nu_5)$}
We are interested in calculating
\be
K(\nu_1,\ldots, \nu_5) \equiv  \int_{\mathcal D_1} x^{-\nu_5} y^{-\nu_4} J(\nu_1,\nu_2,\nu_3,x,y) \;, 
\ee
where the region of integration is given by
\be
\mathcal D_1 =\{ (x,y)\; |\; \sqrt{x}+\sqrt{y}\geq 1,\; x\leq1,\; y\leq 1 \} \;.
\ee
In this domain the power series representation of $\J(\nu_1,\nu_2,\nu_3;x,y)$ is uniformly convergent, and we can use it to rewrite the integral in the following way
\begin{align}
K(\nu_1,& \ldots, \nu_5) = \frac{\sec(\pi\nu_{23})}{8\sqrt\pi \Gamma(\nu_1) \Gamma(\nu_2) \Gamma(\nu_3) \Gamma(3-\nu_{123})} \int_0^1dy\int_{(1-\sqrt{y})^2}^1 d x  \nonumber \\
&\left[ x^{3/2-\nu_{23}} \sum_{n=0}^\infty  a_{n}(\nu_1,\nu_2,\nu_3)\cdot\; x^{n}\;_2F_1\left(\nu_1+n, \tfrac 32-\nu_2+n, 3-\nu_{23}+2n, 1-y \right) \right. \nonumber \\
& \left. -y^{3/2-\nu_{13}} \sum_{n=0}^\infty a_{n}(\tilde\nu_1,\tilde\nu_2,\tilde\nu_3)\cdot\; x^{n}\;_2F_1\left(\tilde\nu_1+n, \tfrac 32-\tilde\nu_2+n, 3-\tilde\nu_{23}+2n, 1-y \right) \right] \;.
\end{align}
Therefore, the basic integral that we want to solve is the integral over the hypergeometric function. For simplicity, let us define 
\begin{align}
& \kappa(\alpha,\beta,a,b,c) \equiv \int_0^1 dy \int_{(1-\sqrt{y})^2}^1 d x\; x^\alpha y^\beta \;_2F_1(a,b,c,1-y) \;.
\end{align}
The integral in $x$ is straightforward, leading to
\begin{align}
\kappa(\alpha,\beta,a,b,c) & = \frac{1}{1+\alpha} \left[\int_0^1 dy \; y^\beta \;_2F_1(a,b,c,1-y) \right. \nonumber \\
& \quad \left. - 2 \int_0^1 dt \; t^{2\beta+1} (1-t)^{2+2\alpha} \;_2F_1(a,b,c,1-t^2) \right] \;,
\end{align}
where in the second integral we did a change of variables $y=t^2$. Both integrals can be expressed in terms of higher order hypergeometric functions. It is not difficult to find
\begin{align}
\kappa (\alpha,\beta,a,b,c) & = \frac{1}{1+\alpha} \left[ \frac{1}{1+\beta}\;_3F_2\left(\begin{array}{c}1,a,b; \\2+\beta,c ;\end{array}1\right) \right. - 2\cdot{}_4F_3\left(\begin{array}{c}a,b,1+\beta,\frac{3}{2}+\beta;
\\1-d ,\frac{5}{2}+\gamma,3+\gamma ;\end{array}1\right)
\nonumber \\
& \left. -2\cdot{}_4F_3\left(\begin{array}{c} c-a,c-b,1+\beta+d,\frac32+\beta+d;
\\1+d,d+\gamma+\frac{5}{2},3+\gamma +d;\end{array}1\right) \right] \;,
\end{align}
where $d=c-a-b$ and $\gamma=\alpha+\beta$. The integration of ${}_2F_1(\cdots,1-y)$ function is straightforward. To integrate ${}_2F_1(\cdots,1-t^2)$ we first have to use~\eqref{eq:zto(1-z)}, expand the hypergeometric functions in power series, integrate in series, then resum the result.

\end{document}